\pdfoutput=1
\documentclass[aps,prb,twocolumn,longbibliography,superscriptaddress,nofootinbib]{revtex4-2}
\usepackage[utf8]{inputenc}
\usepackage[T1]{fontenc}

\usepackage{amsmath}

\usepackage{bm}
\usepackage{bbold}
\usepackage{amssymb}
\usepackage{graphicx}
\usepackage[colorlinks = true, citecolor = blue, urlcolor = blue, linkcolor = blue]{hyperref}
\usepackage{verbatim}
\usepackage{xcolor}
    \definecolor{myred}{RGB}{214, 39, 40}
    \definecolor{mygreen}{RGB}{44, 160, 44}
\usepackage[capitalise]{cleveref}
\usepackage[normalem]{ulem}
\usepackage{dsfont}
\usepackage{todonotes}

\usepackage{accents}

\renewcommand{\th}{\mathrm{th}}
\newcommand{\fb}{\mathrm{fb}}
\newcommand{\ct}{\mathrm{ct}}
\newcommand{\subref}[2]{\hyperref[#1]{\ref*{#1}#2}}

\DeclareMathOperator{\imp}{imp}
\DeclareMathOperator{\BA}{BA}

\begin{document}

\title{Magnetic feedback cooling of a Meissner-levitated micromagnet}

\author{Andrea Marchese}
\thanks{These authors contributed equally to this work.}
\affiliation{CNR-IFN, I-38123, Trento, Italy}
\affiliation{Fondazione Bruno Kessler (FBK), I-38123, Trento, Italy}
\author{Pietro Oreglia}
\thanks{These authors contributed equally to this work.}
\affiliation{CNR -- INO Pitaevskii BEC Center and Dipartimento di Fisica, Universit\`a di Trento, Trento, Italy}
\affiliation{INFN-TIFPA, Trento Institute for Fundamental Physics and Applications, Via Sommarive 14, I-38123 Trento, Italy}
\author{Gianluca Rastelli}
\email{Contact author: gianluca.rastelli@ino.cnr.it}
\affiliation{CNR -- INO Pitaevskii BEC Center and Dipartimento di Fisica, Universit\`a di Trento, Trento, Italy}
\affiliation{INFN-TIFPA, Trento Institute for Fundamental Physics and Applications, Via Sommarive 14, I-38123 Trento, Italy}
\author{Andrea Vinante}
\email{Contact author: andrea.vinante@ifn.cnr.it}
\affiliation{CNR-IFN, I-38123, Trento, Italy}
\affiliation{Fondazione Bruno Kessler (FBK), I-38123, Trento, Italy}

\date{\today}

\begin{abstract}
We cool a librational mode of a levitated micromagnet in a superconducting Meissner trap 
using a magnetic feedback technique. 
To detect the Brownian motion of the micromagnet and to generate the required feedback force, 
we employ a dedicated system of magnetic coils integrated into the trap and positioned close to the micromagnet.
We use two distinct and uncorrelated detection coils to monitor the Brownian motion: 
the first one is employed in the feedback loop and is affected by the squashing effect, 
whereas the second one is not part of the loop. 
We apply feedback cooling at two different pressure values: a high-pressure regime, characterized by a quality factor $Q_{hi-p} \simeq 7.7 \cdot 10^{2}$, 
and a low-pressure regime, characterized by a quality factor $Q_{low-p} \simeq   1.6 \cdot 10^{5}$. 
We analyze our experimental data according to a theoretical model for our specific feedback filter and scheme. 
Estimating  a temperature of $7.9 \pm 1.3~\mathrm{K}$ in absence of feedback, 
 we cool the librational mode down to a temperature of $6 \pm 1~\mathrm{mK}$ at low pressure.
\end{abstract}

\maketitle

%
%
%
%
\section{Introduction}

Quantum nanomechanical \cite{BOOK:Cleland,Poot:2012} 
and optomechanical \cite{Aspelmeyer:2015,BOOK:Bowen} 
systems are massive and spatially extended compared to atoms or molecules.
This makes them ideal platforms for exploring the quantum-classical boundary in macroscopically sized systems, 
enabling precision tests of quantum measurement theory \cite{Hertzberg_Rocheleau_Ndukum_Savva_Clerk_Schwab_2010,Rossi_Mason_Chen_Tsaturyan_Schliesser_2018}, systematic investigations of decoherence mechanisms \cite{youssefi2023squeezed}, 
and the realization of interfaces between different quantum systems \cite{Clerk_Lehnert_Bertet_Petta_Nakamura_2020}. 

In this context, levitated systems \cite{Gonzalez:2021} are playing an increasingly central role, as they feature exceptional isolation from the environment, due to the absence of clamping losses, and large tunability of their properties by external potentials, prospecting quantum control of mechanical motion beyond what is achievable with conventional nanomechanical or optomechanical resonators. 
The long term roadmap of levitodynamics \cite{Gonzalez:2021} includes the realization of largely delocalized nonclassical quantum states, which may enable testing the macroscopic limits of quantum mechanics \cite{Romero2011, Pino2018} and even the quantum nature of gravity \cite{Marletto2017, Bose2017}.

Optical trapping, 
first demonstrated around fifty years ago by Ashkin \cite{Ashkin:1977}, is nowadays the major workhorse in levitodynamics, with some hallmark demonstrations such as motional cooling to the ground state \cite{Delic:2019, Windey:2019, Tebbenjohanns:2021, Magrini:2021, Dania:2025, Troyer:2025} and first investigations of nonclassical states \cite{Rossi:2025}.
However, it is commonly recognized that optical trapping suffers 
from intrinsic limitations, arising from both photon recoil and large internal temperature, the latter leading to unavoidable decoherence channels \cite{Romero2011}. 
For these reasons, several groups are developing 
different strategies based 
on electrical \cite{Dania:2021,Dania2024,Tracy2024} 
or magnetic Paul traps \cite{Bas:2024}, 
magnetic levitation
\cite{Hofer:2023,Vinante:2020PRApplied,Fuchs:2024,Jason:2024}, 
or hybrid platforms where state preparation is engineered via optical trapping and free quantum evolution is performed in the dark with a different method \cite{Bonvin:2024,Nadine:2025}.

%
%
%
%
\begin{figure}[t]
\centering
\hspace{5mm} \includegraphics[width=\linewidth]{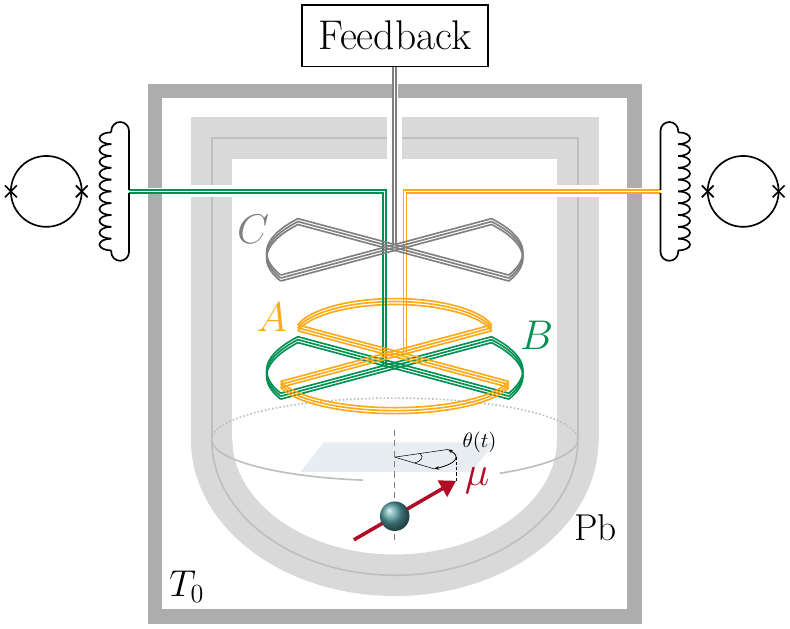}
\caption{
Illustration of the experimental superconducting Meissner trap and geometry of the coils system. 
Coils $A$ and $B$ are used for detection and are inductively coupled to dc-SQUIDs,
whereas coil $C$ is used for feedback cooling (see text). 
The Meissner trap is formed by hemispherical-shaped cup of lead (Pb) with the micromagnet of magnetic moment $\mu$ at the bottom. 
The red arrow indicates the direction of the magnetization. 
The angle $\theta(t)$ describes the angular motion in the plane of the magnetization, parallel to the bottom of the trap.
}
\label{fig:system}
\end{figure}
%
%
%

In order to engineer the quantum state of a levitated mechanical system, a preliminary enabling step is the preparation of a mechanical mode in the quantum ground state. 
Two main strategies can be devised, passive cavity cooling~\cite{Kiesel:2013} or active feedback cooling~\cite{Ashkin:1977}. 
Unlike cavity-cooling, which requires coherent coupling to an optical cavity, either exploiting optomechanical protocols~\cite{Kiesel:2013} or coherent scattering~\cite{Delic:2019, Windey:2019, Delic:2020, Ranfagni:2021, Pontin:2023, Dania:2025, Troyer:2025}, feedback-cooling can be applied to virtually any platform, requiring only sensitive position measurement and a means to apply a feedback force to control position. 
Indeed, feedback-cooling has been demonstrated on a large number of mechanical platforms and is particularly useful in the context of levitodynamics, especially because it allows for cooling of multiple modes. 
In optically trapped nanoparticles, feedback cooling has been implemented using linear~\cite{Ashkin:1977, Li:2011, Conangla:2019, Tebbenjohanns:2019, Tebbenjohanns:2021, Magrini:2021, Kamba:2023} or parametric feedback protocols~\cite{Gieseler:2012, Vovrosh:2017, Penny:2021}. 
Recent advances in detection efficiency have enabled several groups to feedback-cool translational~\cite{Tebbenjohanns:2021, Magrini:2021, Kamba:2023} modes of a levitated nanoparticle to the ground state of motion. Simultaneous cooling of several rototranslational modes~\cite{Li:2020, Kamba:2023} to a few phonons has also been reported, an important ingredient in suppressing nonlinear effects in quantum experiments. 
At the same time, feedback-cooling has been reported in other platforms, namely in Paul traps~\cite{Dania:2021, Bykov:2024} combining optical detection and optical or electrical feedback, in diamagnetically levitated particles~\cite{Slezak:2018, Tian:2024}, in levitated superconducting particles~\cite{Hofer:2023} and in Meissner-levitated ferromagnets~\cite{Timberlake:2024}. 
In the latter case, a ferromagnetic cylinder levitated in a superconducting trap was cooled by combining SQUID-based detection with piezo-driven actuation applied to the trap itself.

Previous works in magnetically levitated systems
reached effective temperatures $T_{\mathrm{eff}}\sim 300$ mK~\cite{Timberlake:2024}, $T_{\mathrm{eff}}\sim 320$ mK~\cite{Tian:2024}, and more recently $T_{\mathrm{eff}}\sim 6$ mK~\cite{Uitenbroek:2026}. All these experiments were performed with ferromagnets or graphite masses of mass $m\sim 10^{-5}-10^{-4}$ kg.

In this work, we report on a feedback cooling scheme for the librational mode of a micromagnet levitating in a superconducting Meissner trap, which relies on purely magnetic torque induced on the suspended particle.
Notably, we cool for the first time a librational mode of a magnetically levitated system down to the millikelvin range. 
Compared to previous attempts, the mass $m\sim10^{-9}$ kg of our magnet is much smaller and we implement a double detection scheme that allows us to witness cooling in the squashed regime.
Furthermore, we outline 
a feasible, though challenging, pathway towards cooling a mode close to the ground state.

The setup is shown in Fig.~\ref{fig:system}. 
We introduce a set of magnetic coils inside the trap, magnetically coupled to the micromagnet.
Two of these coils form two distinct detectors: the motion of the micromagnet induces variations in the magnetic field, which in turn produce changes in the magnetic flux threading the detection coils.
The third coil is part of the feedback loop and functions as the feedback actuator, generating a time-dependent magnetic torque on the micromagnet.
By measuring the power spectral density of the detected signals, we infer cooling of the micromagnet by two or three orders of magnitude under different pressure conditions (i.e., corresponding to different damping regimes of the micromagnet motion).
We analyze the experimental data using a theoretical model of the feedback loop, taking into account both the intended feedback mechanism and a direct magnetic coupling between one detection coil and the actuation coil.

The paper is organized as follows.
In Section \ref{sec:experimental-setup}, we describe the magnetically levitated system and the experimental apparatus.
In Section \ref{sec:theory}, we introduce the theoretical model and the relevant measured quantities, namely the power spectral densities (or simply the power spectra) of the two detectors.
In Section \ref{sec:experimental-results}, we present and analyze the experimental results obtained from the two detectors and compare them with the predictions of the theoretical model.
Section \ref{sec:Temperature} discusses in detail the effective temperature achieved with magnetic feedback cooling as a function of the gain.
In Section \ref{sec:discussion}, we discuss the results and outline a pathway towards approaching ground-state cooling with an improved version of this experiment.
In Section \ref{sec:conclusions}, we draw our conclusions.

%
%
%
%
\section{Experimental setup}
\label{sec:experimental-setup}

A schematic overview of the experimental setup is shown in Fig.~\ref{fig:system}. 

The micromagnet (blue sphere) is taken from an NdFeB alloy powder \cite{MagneticPowder}, magnetized using a 10 T NMR magnet before the experiment. 
The mass of the micromagnet is $m = 1.4 \times 10^{-9}$ kg and radius $R= 35.5 
\, \mu$m. 
The moment of inertia computed from the values of the mass and radius is $\mathcal{I} = 7.06  \times 10^{-19}$ kg m$^2$.

The micromagnet is placed in a superconducting Meissner trap made of Pb with 99.95\% purity. The trap has a cylindrical shape with diameter $5$ mm, and a hemispherical  bottom with radius $2.5$ mm. The whole setup is enclosed in a vacuum chamber that is dipped in liquid helium \cite{Vinante:2020, Ahrens:2025}.
At the operating temperature $T_{\textrm{0}}=4.2$ K, well below the critical temperature of Pb, the magnet spontaneously levitates by the Meissner effect above the bottom of the trap. 
At equilibrium, the magnetic moment lies on the horizontal plane, featuring translational and rotational degrees of freedom. 

In this work, we focus on the rotation in the horizontal plane, parameterized by an angle $\theta$, see Fig.~\ref{fig:system}.
The cylindrical symmetry of the system suggests that the magnet should rotate freely along $\theta$. In practice, as noticed in previous works \cite{Vinante:2020PRApplied, Ahrens:2025}, the symmetry is found to be broken, likely by a residual magnetic field frozen inside the trap. 
The magnet is thus be trapped at an equilibrium angle $\theta_0$  featuring a librational mode around it.

Above the micromagnet, we place two superconducting NbTi coils (A and B), wound on a polymeric holder, that we use for detecting the motion of the magnet via the induced changes of magnetic flux. 
The coils $A$ and $B$ are coupled to two separate dc-SQUIDs, shown in Fig.~\ref{fig:system}, which convert the magnetic flux changes into amplified voltage signals $V_A$ and $V_B$. 
The coils have an eight-shaped design, which maximizes the sensitivity to rotational motion along $\theta$. The coils are placed orthogonal to each other and are located on a plane roughly 2 mm above the micromagnet. Due to symmetry, the magnetic flux $\Phi$ coupled by the magnet into each coil at equilibrium is proportional to $\cos(\theta_0)$ and $\sin(\theta_0)$ for the coils $A$ and $B$, respectively, where we have assumed that $\theta=0$ corresponds to the axis of coil $A$. 
The sensitivities to small flux changes 
is proportional to the derivative $d\Phi/d\theta|_{\theta=\theta_0}$   and is  thus  proportional to $\sin(\theta_0)$ and $\cos(\theta_0)$ respectively.
For small librations driven by thermal motion, relevant to this work, the ratio of the amplitudes of the signals coupled into the coils $A$ and $B$ is thus equal to $\tan(\theta_0)$, assuming the coils are identical.

A third eight-shaped coil (C) is placed above the micromagnet, along with coils A and B, and is used to generate a time-dependent magnetic field to provide feedback to the micromagnet. 
The orientation of coil C was chosen to be orthogonal to coil A, which is used to get the signal for the feedback-cooling circuit. 
This suppresses the crosstalk between the feedback coil and the detection coil. 
Experimentally, we observe that the crosstalk mutual inductance between coils A and C is less than $0.2$ percent 
of that between coils B and C (see Appendix~\ref{app:ringdown_cross_talking}).

The whole system is enclosed in a CryoPerm shield to reduce the external magnetic field (by a factor of roughly 30), and is immersed in helium gas, in thermal contact with a liquid helium bath. The pressure of the helium gas was measured with a Pirani-Penning gauge placed at the top of the measurement probe at room temperature.

Importantly, we used a cryostat with liquid helium to avoid the large vibrational noise typical of dry cryostats employing closed-cycle cryocoolers. In addition, we place the cryostat on an active vibrationally isolated platform, which provides up to $40$ dB attenuation at frequencies higher than 10 Hz.
We performed measurements at two different values of pressure reported in Tab.~\ref{tab:parameters}. 
Other details of the trap are reported in 
Ref.~\cite{Vinante:2020PRApplied,Ahrens:2025,Ahrens:2026}.

%
%
%
%
%
%
%
%
%
%
%
%
\begin{table}[t]
\centering
\begin{tabular}{|c|c|c|c|c|}
\hline
Cryostat temperature            &  High pressure      &  Low pressure     \\
\hline
$T_{\textrm{0}}$  (K) &  $p^{(hi)}$ (mbar)  &  $p^{(lo)}$ (mbar) \\
\hline
$4.2$                  & $1.2 \cdot 10^{-1}$       &  $7 \cdot 10^{-4}$    \\
\hline
\end{tabular}
\caption{
Temperature and 
pressures of the superconducting Meissner trap.
}
\label{tab:parameters}
\end{table}
%
%
%
%
%
%

\begin{figure}[t]
\centering
\includegraphics[width=1.05\linewidth]{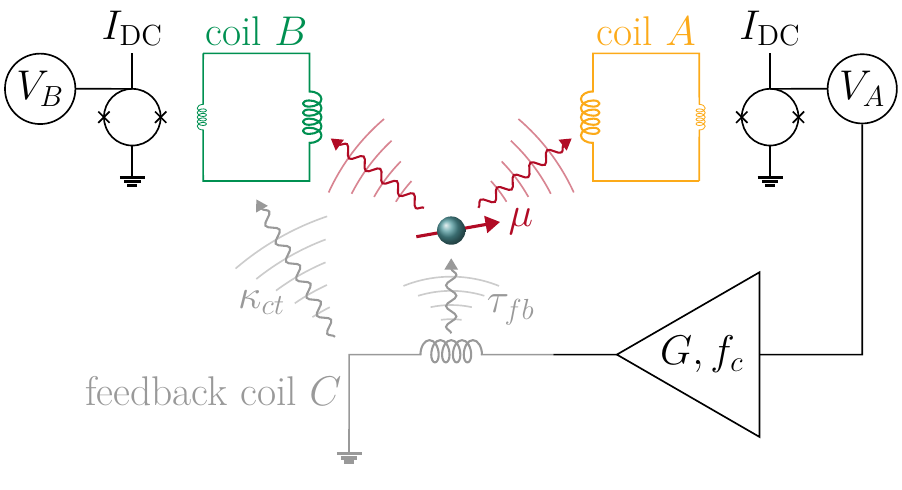}
\caption{ 
Basic conceptual scheme of the feedback loop and the detection mode.
Coils $A$ and $B$ capture the variation of the magnetic flux associated with the motion of the micromagnet.
They are inductively coupled to two dc-SQUIDs, converting the time-dependent magnetic flux variation into voltages, 
$V_A(t)$ and $V_B(t)$.
The signal from detector $A$ is sent to the feedback loop, where, within a given frequency range,
it is processed through a filter function
with a low-frequency cutoff $f_c$, amplified by a gain $G$, 
and
converted into a time-dependent magnetic field via coil $C$.
This produces an effective feedback torque $\tau_{\fb}$ acting on the micromagnet.
The magnetic field produced by the coil $C$ is also detected by the coil $B$ (cross-talk effect). 
}
\label{fig:scheme}
\end{figure}

In Fig.~\ref{fig:scheme} we report the conceptual scheme.
For detection, coils $A$ and $B$ are sensitive to variations of the magnetic field caused by the stochastic motion of the micromagnet.
Each of them is inductively coupled to a dc-SQUID (low $T_c$ two-stage current sensor from Magnicon), 
which converts the time-dependent magnetic flux variations into voltage. The SQUIDs are controlled by a two-channel electronics (Magnicon XXF-1).

Hereafter, we refer to the ensemble of each coil with its respectively coupled dc-SQUID as the detector $A$ and the detector $B$, 
enabling the conversion of the angular motion in time $\theta(t)$ into voltage signals, $V_A(t)$ and $V_B(t)$  respectively.

The feedback signal is taken from the output of the detector $A$, passed through a one-pole low-pass filter (Stanford Research SR560) with cut-off frequency $f_c$ and tunable gain $G$ before being fed to the coil $C$.

%
%
%
%
%
%
%

%
%
%
%
\section{Theoretical model}
\label{sec:theory}

\subsection{Langevin equation with feedback}

Hereafter, we focus on the angular motion of the micromagnet, described by the angle $\theta(t)$ in the $xy$ plane, with a rotation axis perpendicular to the magnetic moment $\mu$, see Fig.~\ref{fig:system}. 
In a perfectly symmetric Meissner trap with an hemispherical bottom, and with the micromagnet located at the center of the trap, this motion would correspond to that of a free rotor.
As explained above, however, we experimentally observe that this symmetry is broken,
and the micromagnet exhibits librational motion around an equilibrium position. 
For small oscillation amplitudes, this motion corresponds to that of a harmonic oscillator with angular frequency $\omega_0 = 2\pi f_0$ and frequency $f_0$.

The dynamics of the stochastic small harmonic oscillations of the angle $\theta$ under applied feedback
is described by the  Langevin equation 
%
%
%
%
%
%
%
\begin{equation}
\frac{d^2\theta(t)}{dt^2} 
= 
- \omega_0^2 \theta(t) - \Gamma_0 \frac{d\theta(t)}{dt} + \delta \tau_{\mathrm{th}}(t) + \tau_{\mathrm{fb}}(t)  
\label{eq:Langevin}
\end{equation}
%
%
%
%
%
%
with  $\Gamma_0$ the pressure-dependent  damping coefficient, 
$\delta \tau_{\th}(t)$ the thermal noise with the property 
%
%
%
%
%
%
%
\begin{equation}
\langle \delta \tau_{\mathrm{th}}(t) \delta \tau_{\mathrm{th}}(t')  \rangle 
=  
S_{\mathrm{th}} \, \delta(t-t') \, , \quad  \mbox{with}  \quad S_{\mathrm{th}} = \frac{ 2  \Gamma_0 k_B T}{ \mathcal{I} }
\label{eq:thermal_noise}
\end{equation}
%
%
%
%
%
 %
where 
$T$ is the temperature in the absence of feedback, which we refer to as the zero-gain temperature.
$\mathcal{I}$ is the moment of inertia associated with the libration motion  ($k_B$  is the Boltzmann constant) and 
$ \tau_{\mathrm{fb}}(t)  $ the feedback torque as detailed below.
Notice that $\delta \tau_{\mathrm{th}}(t) $ and 
$\tau_{\mathrm{fb}}(t)  $ are scaled torques normalized with the moment of inertia.

\subsection{The feedback loop}

As shown in Fig.~\ref{fig:scheme}, 
the output voltage $V_A(t)$ from the dc-SQUID inductively coupled to coil $A$ 
is frequency-filtered using a filter function and amplified by a specified gain before being converted into a time-dependent magnetic field via coil $C$. 
This results in an effective feedback torque $\tau_{\mathrm{fb}}$ acting on the micromagnet.
Taking into account the intrinsic noise of the dc-SQUID, the torque acting on the micromagnet is given by 
%
%
%
%
%
%
%
%
%
%
\begin{equation}
\tau_{\mathrm{fb}}(t)
 = 
 \int^{t}_{-\infty} \!\!\!\!\! dt' 
 K_{\mathrm{fb}}\left( t-t' \right) 
 \left[ 
 \theta(t')  + \delta\theta_{A}(t') 
 \right] \,
 \label{eq:fb_function}   
\end{equation}
%
%
%
%
%
with $ \delta\theta_{A}(t)$ the angular noise associated with the output of the dc-SQUID coupled to coil $A$, as detailed below. 
In our experiment, we use the following feedback function   
%
%
%
%
%
%
%
%
%
\begin{equation}
K_{\mathrm{fb}}(\omega) = \frac{\kappa_0}{\omega_c - i \omega} \,,
\end{equation}
%
%
%
%
%
with $\omega_c$ a low-frequency cutoff. 
In our apparatus, we have $\omega_c/(2\pi)=30$ Hz. 
The parameter $\kappa_0$ encapsulates the total gain and all proportional constants involved 
in the filtering loop that links the voltage output from the dc-SQUID coupled to the coil $A$ 
to the injected signal, namely the magnetic field generated by the coil $C$. 
It also incorporates the geometrical form factor governing the coupling between the magnetic field 
produced by the coil $C$ and the micromagnet.

It is useful to make a connection with the cold damping filter in feedback cooling schemes~\cite{Poggio:2007, Penny:2021}.
This is possible if we define the scaled gain of the theoretical model  as the dimensionless quantity 
%
%
%
%
%
%
%
\begin{equation}
g = \frac{\kappa_0}{\Gamma_0 \omega_0^2} 
\label{eq:g}
\,.
\end{equation}
%
%
%
%
%
%
Using $g$, then we note that in the limit of $\omega_c\ll \omega_0$, we can approximate the feedback  function  as $K_{\mathrm{fb}}(\omega)  \approx i g \Gamma_0 \omega_0$,
since the relevant frequency range is the one associated with the resonant peak of the angular motion $(\omega \approx \omega_0)$.
 In the underdamped regime 
 $(\Gamma_0 \ll \omega_0)$, the quantity  
$\tau_{\mathrm{fb}}(\omega) = i g \Gamma_0 \omega_0 \theta(\omega)$  
approximately represents the quantity 
$( -g\Gamma_0 d\theta/dt)$ in time domain, 
namely 
it is equivalent to the cold damping filter of feedback cooling schemes \cite{Poggio:2007,Penny:2021}.

\subsection{The detection signal}

Assuming that all filtering and amplification functions (not shown in the basic scheme of Fig.~\ref{fig:scheme}
) are linear, we can write  for the detector $A$
%
%
%
%
%
%
%
\begin{equation}
V_{A}(t)  = a_A \theta(t) + \delta V_A(t)  = a_A \Big[   \theta(t) + \delta \theta_A(t)  \Big]
\label{eq:V_A}
\end{equation}
%
%
%
%
%
with $V_{A}(t)$ the output voltage
and  $a_A$ corresponding to the global calibration factor for the detector $A$.
$\delta V_A(t)$ is the intrinsic noise of the SQUID that sets the corresponding noise of the measurement, namely 
$\delta\theta_{A} = \delta V_A(t)/a_A$.

A similar relation holds for the detector $B$.
However, in our experimental setup, by geometry, the coil of the detector $B$ is also sensitive to the  
magnetic field generated by the coil $C$ so that the output voltage of the detector $B$ is given by 
%
%
%
%
%
%
%
\begin{equation}
\frac{ V_{B}(t) }{a_B}  
=
\theta(t)  
+ \delta\theta_B(t) 
+ \int^{t}_{t_0} \!\!\! dt' K_{\mathrm{ct}} \left( t-t' \right) \left[ \theta(t')  + \delta\theta_{A}(t) \right] 
\, .
\label{eq:V_B}
\end{equation}
%
%
%
%
%
%
By the geometry of the system, we assume that $a_A\neq a_B$ since the two coils $A$ and $B$ enclose a different magnetic flux associated with the micromagnet and, furthermore, the two dc-SQUIDs are not identical.

The third term on the right side of Eq.~\ref{eq:V_B} 
 is proportional to the magnetic field signal produced by the coil $C$ 
 which depends on the gain $g$.
 However,  the cross-talk function $K_{\mathrm{ct}}(t) $
also depends on the mutual inductance between the coil $C$ and the coil $B$  which is different from the inductive coupling  
between $C$ and the micromagnet; therefore, we set
%
%
%
%
%
%
%
%
\begin{equation}
K_{\mathrm{ct}} (\omega) = \frac{ \kappa_{\mathrm{ct}}  }{\omega_c - i  \omega} \, ,
\label{ed:ct}
\end{equation}
%
%
%
%
%
where  the geometrical factors of the mutual inductance between the coil $C$ and the coil $B$  as well as all other proportional constants of the feedback loop, including the gain,  are 
incorporated in the parameter $\kappa_{\mathrm{ct}} $. 
As before, it is useful to parametrize $\kappa_{\mathrm{ct}} $ using the dimensionless constant $\eta$ 
%
%
%
%
%
%
%
%
\begin{equation}
\frac{ \kappa_{\mathrm{ct}}  }{\omega_c} = g \, \eta
\, ,
\end{equation}
%
%
%
%
%
%
where we have made the dependence of the gain $g$ explicit.

As explained before, the orientation of the coil $B$ is exactly the same as that of the coil $C$ 
whereas the coil $A$ has its axis orthogonal to that of the feedback coil $C$,
see Fig.~\ref{fig:system}. 
For this reason, the cross-talk between the coils $C$ and $A$ is much smaller, and we neglect it 
(see Appendix \ref{app:ringdown_cross_talking}).

\subsection{Power spectra}

In this section, we report the theoretically computed power spectra of the different variables: $\theta(t), V_A(t), V_B(t)$.
In particular, the power spectra of $V_A(t)$ and $V_B(t)$   are used to fit the experimental data in the next section.
Further details of the calculation are reported in Appendix \ref{app:theory}.

\subsubsection{Power spectra of the detectors' noise}

We introduce the power spectra of the intrinsic  noise at the output of the dc-SQUIDs, which are coupled to the coil $A$ and to  the coil $B$ 
%
%
%
%
%
%
%
%
\begin{align}
S_{\delta V}^{(A)}(\omega) 
&=
\int^{+\infty}_{-\infty} \!\!\!\!\!
d\tau \,\,
\langle \delta V_A(\tau) \delta V_A(0) \rangle \, e^{i\omega\tau} \,, \\
S_{\delta V}^{(B)}(\omega) 
& =
 \int^{+\infty}_{-\infty}  \!\!\!\!\!
 d\tau\,\,
 \langle \delta V_B(\tau) \delta V_B(0) \rangle \, e^{i\omega\tau} \,.
\end{align}
%
%
%
Experimentally, we observe that 
in the frequency range from  a few Hz up to several hundred Hz, $S_{\delta V}^{(A)}(\omega)$ and 
$S_{\delta V}^{(B)}(\omega)$  are almost constant and independent of the frequency 
so we set the constant $S_{\delta V}^{(A)}(\omega)  \equiv S_{\delta A} $ and $S_{\delta V}^{(B)}(\omega)  \equiv S_{\delta B} $.

\subsubsection{Power spectrum of $\theta\left( t \right) $}

From  Eq.~(\ref{eq:Langevin}) and  Eq.~(\ref{eq:fb_function}), 
 the relation  in frequency space that links the angle to the thermal noise and the detection noise reads 
%
%
%
%
%
%
%
%
\begin{equation}
\theta\left( \omega \right) 
=
\chi\left( \omega \right) 
 \left[ 
 \delta\tau_{\mathrm{th}} \left( \omega \right) + K_{\mathrm{fb}}(\omega) \delta \theta_A\left( \omega \right) 
 \right]
\, ,
\label{eq:Langevin_omega}
\end{equation}
%
%
%
%
%
%
%
%
where we introduced the response function, the generalized susceptibility including the feedback function, for the harmonic oscillatory motion 
%
%
%
%
%
%
%
%
\begin{equation}
\chi\left( \omega \right) 
=
 \frac{1}{\omega_0^2- \omega^2 - i \Gamma_0 \omega - K_{\mathrm{fb}}(\omega)} 
\, .
\label{eq:response_f}
\end{equation}
%
%
%
%
%
Using Eq.~(\ref{eq:Langevin_omega})    we obtain the power spectrum $S_{\theta}\left( \omega \right) $ of the variable $\theta(t)$ in the presence of the feedback 
%
%
%
%
%
%
%
%
\begin{align}
S_{\theta}\left( \omega \right) 
&= 
 \int^{+\infty}_{-\infty} \!\!\!\!\!
 d\tau \,\,
  \langle \theta(\tau) \theta(0) \rangle \, e^{i\omega\tau} \nonumber \\
&
= 
 {\left| \chi \left( \omega \right) \right|}^2 \left[  S_{\mathrm{th}} +    \frac{1}{a_A^2}  {\left| K_{\mathrm{fb}} \left( \omega \right) \right|}^2     S_{\delta A}  \right]
 \label{eq:S_theta}
\end{align}
%
%
%
%
%
where $S_{\th}$ is the thermal noise as given by Eq.~(\ref{eq:thermal_noise})
whereas $S_{\delta A} $ is the noise of the detector $A$.
We observe that, beyond the thermal noise, the stochastic dynamics of the angle is also affected by the noise of the detector $A$  used 
in the feedback loop, filtered by the feedback function.

%
%
%
%
%
%
%
%
%
%
%
%
%
%
%
%

\subsubsection{Power spectrum of the detector $A$}

From the relation  Eq.~(\ref{eq:V_A}) in frequency space and using  Eq.~(\ref{eq:Langevin_omega}) we compute the power spectrum for the output signal $V_A(t)$ 
%
%
%
%
%
%
%
%
\begin{align}
S_{A}\left( \omega \right) 
&
=
 \int^{+\infty}_{-\infty} \!\!\!\!\!
 d\tau \,\,
 \langle V_A(\tau) V_A(0) \rangle e^{i\omega\tau}   \nonumber \\
&
= a_A^2   \, 
 S_{\mathrm{th}} {\left| \chi \left( \omega \right) \right|}^2  + |1+K_{\mathrm{fb}}\left( \omega \right) \chi\left( \omega \right)|^2 S_{\delta A} \; . 
  \label{eq:V_A_omega}
\end{align}
%
%
%
%
%
Compared to Eq.~\eqref{eq:S_theta},  
$S_{A}\left( \omega \right)$ contains an extra term, which is related to the fact that the noise of detector 
$A$ is re-injected into the system and affects the stochastic dynamics of $\theta(t)$ (see Eq.~\eqref{eq:Langevin} and Eq.~\eqref{eq:fb_function})
leading to the squashing effect in the power spectrum of $V_A(t)$: 
by increasing the gain $g$, the spectrum at some point no longer shows a resonant peak, but instead becomes depleted below the noise level~\cite{Poggio:2007}.

\subsubsection{Power spectrum of the detector $B$}

From the relation  Eq.~(\ref{eq:V_B}) in frequency space, we compute the power spectrum for the output signal $V_B(t)$ 
%
%
%
%
%
%
%
\begin{align}
&
S_{B}\left( \omega \right) 
= 
\int^{+\infty}_{-\infty} \!\!\!\!\!
d\tau \,\, 
\langle V_B(\tau) V_B(0) \rangle e^{i\omega\tau}   
\nonumber \\
&
=
a_B^2 \,
{\left| 
1 + K_{\ct}\left( \omega \right) 
\right|}^2
{\left| 
\chi \left( \omega \right) 
\right|}^2 
S_{\th} 
+
S_{\delta B} 
\nonumber \\
&
+
\,
{\Big| 
\big(  1 + K_{\ct}\left( \omega \right) \big)
\chi \left( \omega \right)  K_{\fb}\left( \omega \right) +  K_{\ct}\left( \omega \right)  
\Big|}^2 
\left( \frac{a_B^2}{a_A^2} \right) S_{\delta A}  
\label{eq:V_B_omega}
\, .
\end{align}
%
%
%
%
%
%
%
%
%
%
%
It is useful to consider the expression for the power spectrum of the detector $B$ in the absence of cross-talk, $\kappa_{\ct}=0$ 
($K_{\ct}(\omega)=0$). We find
%
%
%
%
%
%
%
%
%
%
\begin{equation}
S_{B}^{(\kappa_{\ct}=0)} \left( \omega \right) 
=
a_B^2  S_{\theta}\left( \omega \right)
+
S_{\delta B} 
\, .
\end{equation}
%
%
%
%
%
%
%
%
%
%
%
%
Comparing $S_{B}^{(\kappa_{\ct}=0)}\left( \omega \right) $ with the expression $S_{B}\left( \omega \right) $ 
we observe that the presence of cross-talk leads to a further filtering of the signal, via the term 
$(1 + K_{\ct}\left( \omega \right))$ and an additional contribution related to the noise of the detector A 
(see also Appendix \ref{app:theory}).
The cross-talk yields an asymmetric lineshape of the resonant peak in the power spectrum, namely an increase in spectral weight at frequencies below the resonance frequency.

%
%
%
%
\section{Experimental results}
\label{sec:experimental-results}

Before presenting the experimental spectra for the angular Brownian motion of the two detectors, we first discuss the measurement noise.

The detectors' noise spectra are extracted from the flat background (plateau) of the spectrum far from the resonance peak.
For the two detectors, we have
%
%
%
%
%
%
%
%
\begin{align}
S_{\delta A} &=\left(  1.01 \pm 0.01 \right)  \cdot 10^{-11}  \,\,\,  \mbox{V}^2/\mbox{Hz} \, ,\\
S_{\delta B} &= \left( 0.48 \pm 0.01 \right) \cdot 10^{-11}  \,\,\,    \mbox{V}^2/\mbox{Hz}  \, .
\end{align}
%
%
%
%
%
%
%
%
We experimentally find that $S_{\delta A}$ and $S_{\delta B}$ are independent of the pressure.

\subsection{Power spectra at high pressure without feedback}
\label{subsec:PSD_high_g0}

\begin{figure}[t]
\flushleft{(a)} \\    
\includegraphics[width = 0.9\linewidth]{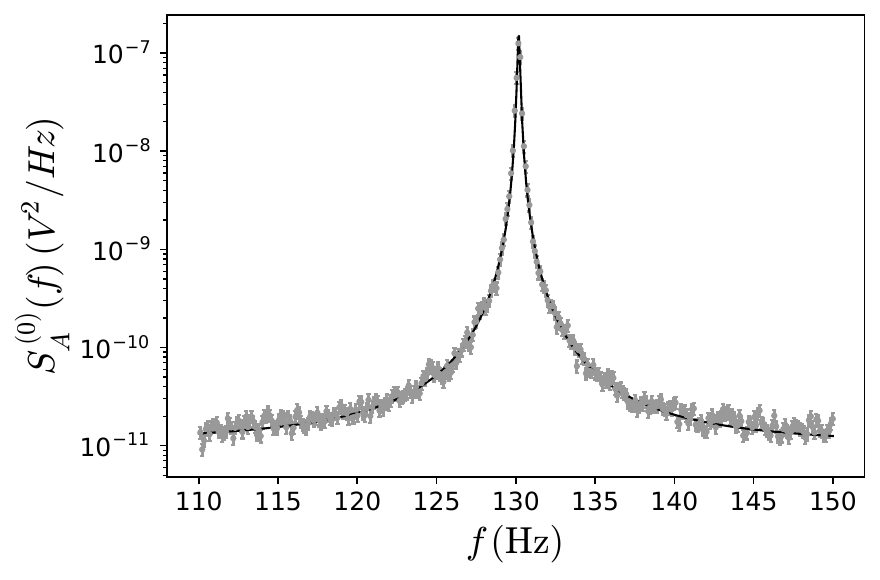}\\
\flushleft{(b)} \\
\includegraphics[width = 0.9\linewidth]{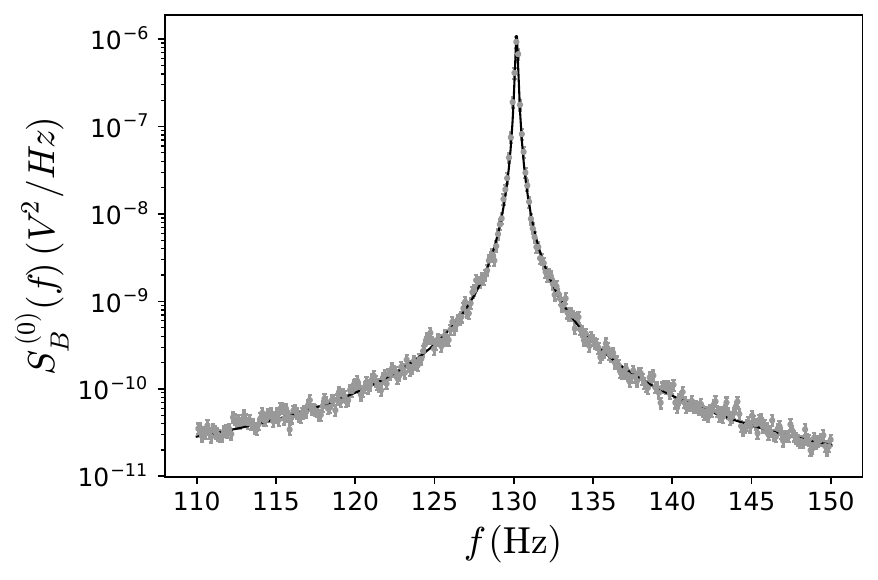}
    \caption{
 Comparison at zero gain ($g=0$, no feedback) between the theoretical fit (lines) and the experimental spectra (points) 
 at the resonance peak of the angular Brownian motion, at high pressure, $p^{(hi)}$.
(a) Power spectrum $S_A^{(0)}\left(f\right)$ for detector A. The line corresponds to the theoretical formula, Eq.~(\ref{eq:S_A_g=0}).
(b) Power spectrum $S_B^{(0)}\left(f\right)$ for detector B. The line corresponds to the theoretical formula, Eq.~(\ref{eq:S_B_g=0}).
Parameters are listed in Table~\ref{tab:fitting_g=0}.
    }
    \label{fig:spectra_g0}
\end{figure}

For the high-pressure case, the power spectra of the SQUID voltages for detectors $A$ and $B$ at zero gain ($g=0$) are shown in Fig.~\ref{fig:spectra_g0}.
For $g=0$, the power spectra of the two detectors reduce to the following expressions
%
%
%
%
%
%
%
%
%
%
%
%
%
\begin{align}
S_{A}^{(0)}\left( \omega \right) 
&=
a_A^2 \,
{\left| 
\chi_0 \left( \omega \right) 
\right|}^2 
S_{\th} 
+
S_{\delta A} 
\, ,
\label{eq:S_A_g=0}
\\
S_{B}^{(0)}\left( \omega \right) 
&=
a_B^2 \,
{\left| 
\chi_0 \left( \omega \right) 
\right|}^2 
S_{\th} 
+
S_{\delta B} 
\label{eq:S_B_g=0}
\, ,
\end{align}
%
%
%
%
%
%
%
%
%
%
%
with $\chi_0 \left( \omega \right) $ the susceptibility for $g=0$.

%
%
%
%
%
%
%
%
%

We can safely assume that the system is thermalized in the high-pressure regime in the absence of feedback, so that the temperature of the angular Brownian motion at zero gain can be taken as $T_{\textrm{0}}$ (see Table~\ref{tab:parameters}). 
Accordingly, we use Eq.~\eqref{eq:S_A_g=0} and Eq.~\eqref{eq:S_B_g=0} to fit the experimental data reported in Fig.~\ref{fig:spectra_g0},
with the fitting parameters given by the resonance frequency $f_0$, the damping coefficient $\Gamma_0$, and the two calibration factors $a_A$ and $a_B$. 
The fitting results are reported in Table~\ref{tab:fitting_g=0}.

%
%
%
%
%
%
%
\begin{table}[b]
\centering
\begin{tabular}{|c|c|}
\hline
\multicolumn{2}{|c|}{
High pressure} \\
\hline
  detector $A$                               &    detector $B$ \\
\hline
   $a_A= 26 \pm 1$  V     &  $a_B = 70 \pm 2 $      V  \\
\hline
$f_{0,A} = 130.182 \pm 0.008$  Hz      &   $f_{0,B}= 130.179 \pm 0.008$  Hz   \\ 
\hline
$\Gamma_{0,A}/(2\pi) = 0.17 \pm 0.01 $   Hz      &   $\Gamma_{0,B}/(2\pi) = 0.17 \pm 0.01 $   Hz   \\ 
\hline
\end{tabular}
\caption{
Results for the fitting parameters at zero gain $g=0$ (no feedback) for the two detectors $A$ and $B$ at high pressure  $p^{(hi)}$.  
$a_A$  and $a_B$ are   the calibration factors,  $f_{0,A} $  and  $f_{0,B} $  the resonance frequencies, 
$\Gamma_{0,A}$   and $\Gamma_{0,B}$   the damping coefficients.}
\label{tab:fitting_g=0}
\end{table}
%
%
%
%
%

From Table~\ref{tab:fitting_g=0} we observe that the two frequencies and the two damping coefficients are consistent within the statistical uncertainty, as expected given that the two channels are measuring the same signal.
The damping rate corresponds to a quality factor $Q_{hi-p} \simeq 766$.

In contrast, the two calibration coefficients differ substantially, indicating that detector $B$ is more strongly coupled to the micromagnet than detector $A$. 
We can estimate the equilibrium angle as $\theta_0 = \arctan(a_A/a_B) \simeq 20 \pm 1$ deg.
We recall, as reference, that the equilibrium angle is zero when the magnetic dipole is parallel to the axis of the coil A, namely perpendicular to the axis of the coil B.

\subsection{Power spectra at high pressure with feedback}

Hereafter, we keep the parameters reported in Table~\ref{tab:fitting_g=0} fixed, as they were determined in the high-pressure case without feedback, in Sec.~\ref{subsec:PSD_high_g0}. 

\begin{figure}[t]
    \centering
\flushleft{(a)} \\      
\includegraphics[width = 0.9\linewidth]{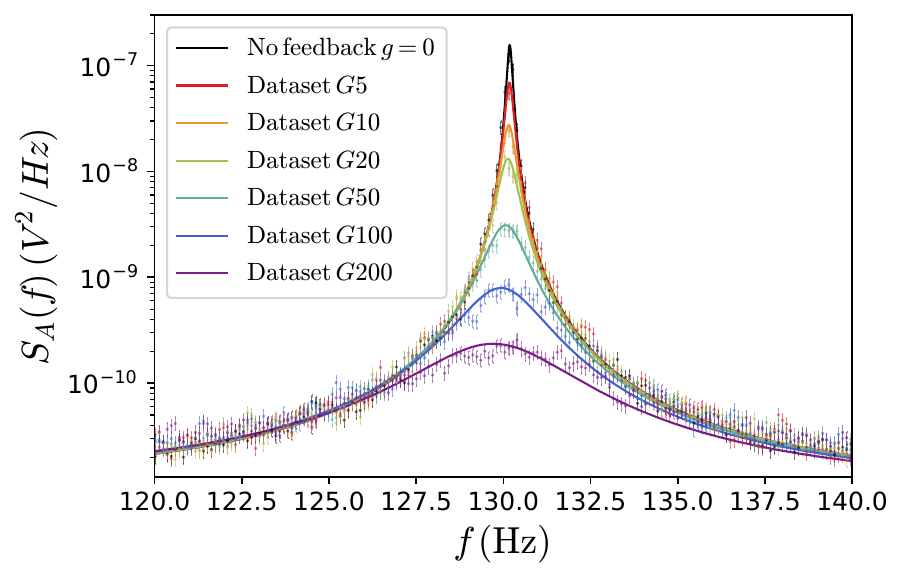}\\
\flushleft{(b)} \\  
\includegraphics[width = 0.9\linewidth]{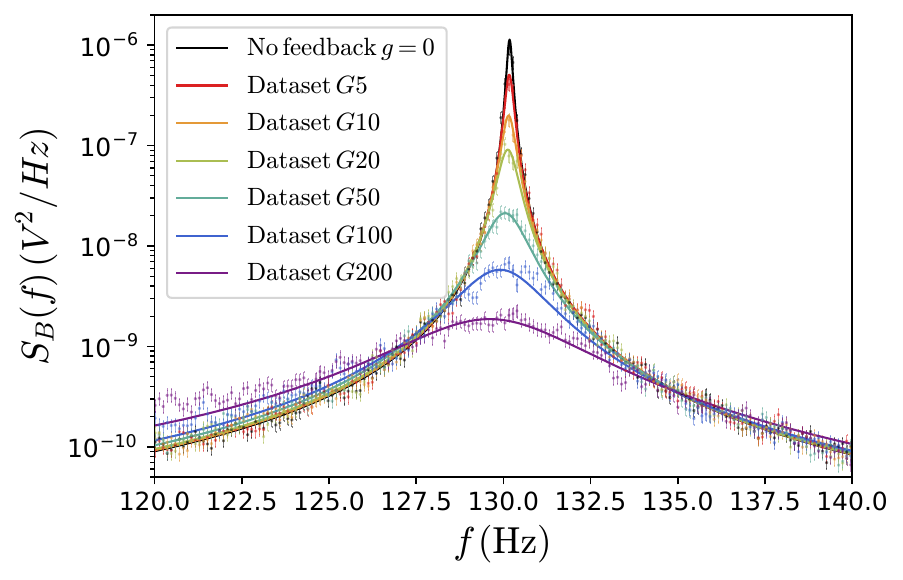}
 \caption{
  Comparison with feedback loop   between the theoretical fit (lines) and the experimental spectra (points) 
 at the resonance peak of the angular Brownian motion, at high pressure, $p^{(hi)}$.
 {\bf (a}) 
 Power spectrum $S_A\left(f\right)$ for the detector A.
 The lines are the theoretical formula Eq.~(\ref{eq:V_A_omega}).
 {\bf (b}) 
 Power spectrum $S_B\left(f\right)$ for the detector B.
 The lines are the theoretical formula Eq.~(\ref{eq:V_B_omega}).}
    \label{fig:spectra_high_P}
\end{figure}

In Fig.~\ref{fig:spectra_high_P}  we report the experimental spectra for the case of high pressure $p^{(hi)}$ 
applying the feedback loop $(g\neq 0)$ for the two detectors, $A$ and $B$. 
The solid curves are the fitting curves.
We tuned the feedback-loop gain by changing the experimental gain $G$ of the feedback amplifier. 
Different curves are labeled by a different value of $G$ in the figures.

In Fig.~\subref{fig:spectra_high_P}{a}, for the detector $A$ in the feedback loop, 
we used the formula  Eq.~(\ref{eq:V_A_omega}) using only  $g$, the scaled gain, 
as a single fitting parameter for the different sets of data at different $G$. 

In Fig.~\subref{fig:spectra_high_P}{b}, for the detector $B$, 
we used the formula  Eq.~(\ref{eq:V_B_omega}) using  $g$, the scaled  gain, 
and the cross-talk $\eta$ 
as two fitting parameters for the different sets of data at different $G$.

%
%
%
%
%
%
%
%
%

In  Fig.~\ref{fig:spectra_high_P}  
we observe a reduction of the areas under the resonance curves with increasing feedback gain, 
which clearly indicates a reduction in the effective temperature of the angular Brownian motion.
Furthermore, we observe good agreement between the theoretical curves and the experimental data.
Considering that only one parameter is used to fit the spectrum $S_A(f)$, namely the scaled  gain $g$, 
whereas two parameters are used to fit the spectrum $S_B(f)$, namely $g$ and the cross-talk constant $\eta$, 
the overall consistency between theory and experiment is remarkable.

For the detector $B$, in Fig.~\subref{fig:spectra_high_P}{b}, we emphasize the slight asymmetry that we anticipated and which arises from the cross-talk effect: the detection coil $B$ also senses the direct signal from the feedback coil $C$ 
and 
this leads to an increase in the spectral weight at frequencies lower than the resonance frequency.

\begin{figure}[t]
 \includegraphics[width =0.9 \linewidth]{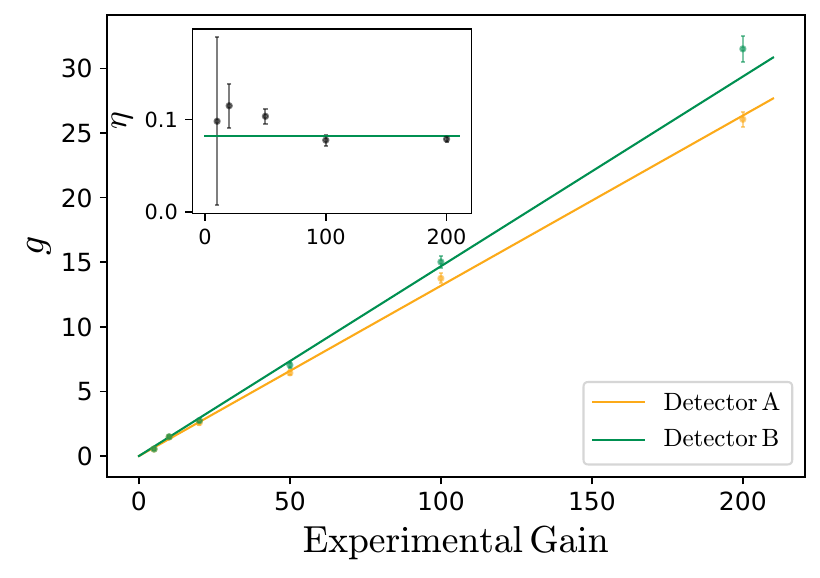}
 \caption{
 For the high-pressure data, $p^{(hi)}$, 
 the fitting parameter $g$ (scaled gain), together with its associated error, is shown for the two detectors across the different data sets acquired at varying experimental gains.
The inset shows the results for the fitting parameter $\eta$, representing the cross-talk coefficient, for detector $B$.
}
    \label{fig:g_high_P}
\end{figure}

In Fig.~\ref{fig:g_high_P}, we report the results of the fit for the parameter $g$ for the two detectors for the different data sets at varying experimental gains.
A straightforward linear dependence is expected and is confirmed by the data.
More importantly, 
within the statistical error, we obtain the same $g$ curve for both detectors
, once again demonstrating the consistency of our theoretical model.
The inset of Fig.~\ref{fig:g_high_P} shows the behavior of $\eta$, the crosstalk parameter used in fitting the data for detector $B$, for the different data sets at varying experimental gains.

%
%
%
%
%
%
%
%
%

\subsection{Power spectra at low pressure with feedback}
\label{subsec:PSD_low}

Lowering the pressure by three orders of magnitude strongly reduces the damping rate of the levitated micromagnet. 
However, we remain in a regime where the damping is still primarily determined by gas friction.
At the low-pressure value $p^{(lo)}$, the angular libration motion 
features a higher quality factor of the order $Q_{low-p} \sim 10^{5}$, in the absence of the feedback loop.
Due to the narrow linewidth, the damping coefficient is more conveniently extracted from independent ring-down measurements, yielding 
$\tilde{\Gamma}_0/(2\pi) =  (8.4 \pm 0.3) \times 10^{-4}$ Hz 
(see Appendix \ref{app:ringdown_cross_talking}). 
Note that hereafter we use the symbols with tilde to indicate the values at low pressure, if they are or can be different from those at high pressure.

For the low-pressure data, in order to achieve a meaningful comparison between the experimental results and 
the theoretical curves, we needed to introduce an additional fitting parameter, 
defined as the product of the zero-gain temperature and the calibration factors, namely
$c_A = \tilde{a}_A^2 T $ and $c_B =  \tilde{a}_B^2 T$. 
This means that we allow for possible variations of the calibration parameters $\tilde{a}_A$ and $\tilde{a}_B$ or the zero-gain temperature of the mechanical torque noise $T$ with respect to the high pressure values $a_A$, $a_B$ and $T_0$.
The reason for this choice is that, in contrast with the high-pressure case in which we can safely assume that the torque noise is thermal
(the cryostat temperature is the zero-gain temperature) 
for low damping, the thermal noise is more than two orders of magnitude lower, so we may in principle have additional small noise contributions (e.g. from vibrational noise).
Furthermore, while we should expect the coefficients $a_A$ and $a_B$ not to change at low pressure, we find a posteriori that this is not the case.
Below, we discuss an explanation.

To determine the resonance frequencies for the two data sets at low pressure, we proceeded as follows.
For the experimental curves corresponding to the lowest gain (dataset $G5$), we used as fitting parameters the resonance frequency $f_0$, 
the scaled gain $g$, and the coefficient $c_A$ for detector $A$.
For detector $B$, in addition to these three parameters $(f_0, g, c_B)$, we also included $\eta$, the crosstalk coefficient, as a fourth fitting parameter.
This procedure allows us to determine the resonance frequencies as
$\tilde{f}_{0,A} = 130.468 \pm 0.008$ Hz and $\tilde{f}_{0,B} = 130.45 \pm 0.02$ Hz.
%
These values yield a quality factor $Q \simeq 156000$.
These values for the two frequencies are slightly larger than the resonance frequency observed at high pressure, 
which we attribute to a possible change in the residual field which determines the librational confinement of the angle $\theta$. 
As we shall see later, this assumption is confirmed by a slight change of the calibration factors (related to the couplings to the two coil detectors) which points to a slight change of the equilibrium angle $\theta_0$.

After that, we keep the two frequencies fixed. For fitting the dataset at larger experimental gains 
 (datasets $G10$, $G20$, $G50$, $G100$, $G200$), we use $(g, c_A)$ as fitting parameters 
 for detector $A$ and $(g, c_B,\eta)$  as fitting parameters 
 for detector $B$.

%
%
%
%
\begin{figure}[t]
    \centering
\flushleft{(a)} \\      
\includegraphics[width = 0.9\linewidth]{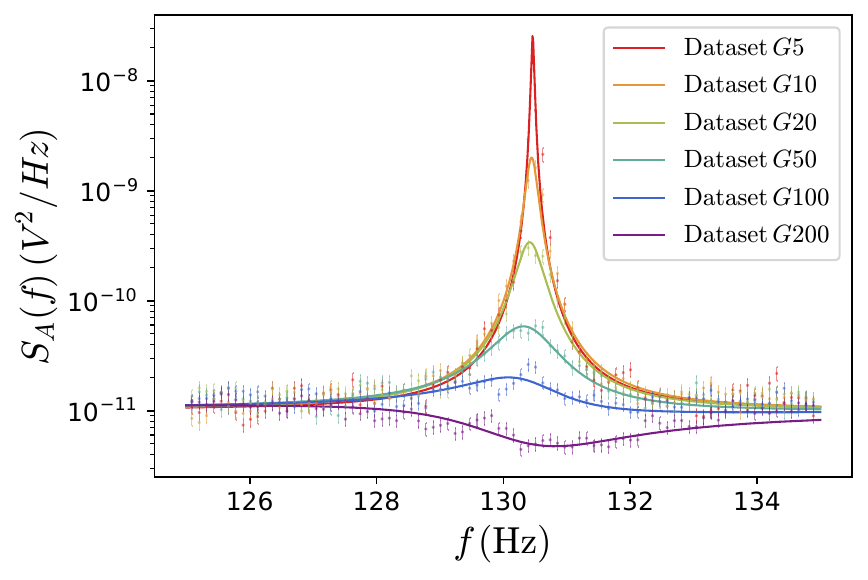}\\
\flushleft{(b)} \\  
\includegraphics[width = 0.9\linewidth]{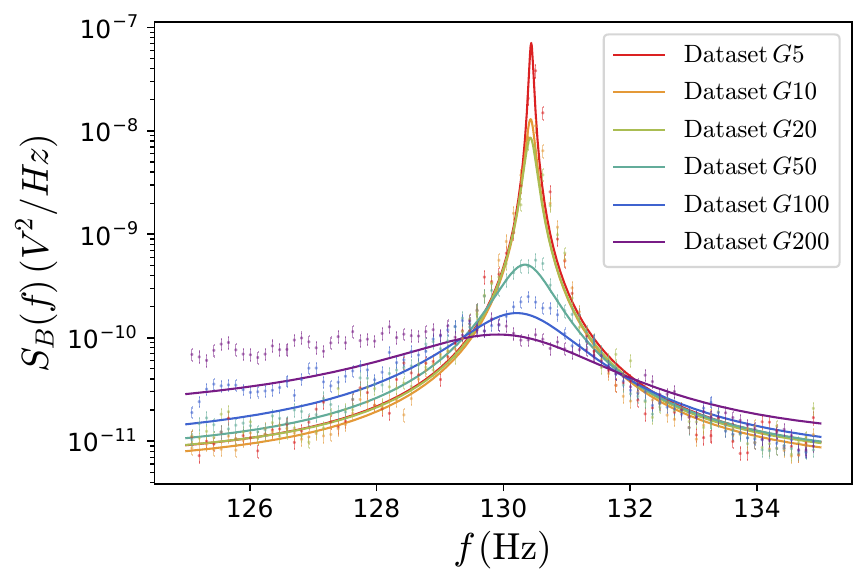}\\
 \caption{
 Comparison with feedback loop between the theoretical fit (lines) and the experimental spectra (points) at
the resonance peak of the angular Brownian motion, at low
pressure, $p^{(lo)}$.
 {\bf (a}) 
 Power spectrum $S_A\left(f\right)$ for the detector A.
 The lines are the theoretical formula Eq.~(\ref{eq:V_A_omega}).
 {\bf (b}) 
 Power spectrum $S_B\left(f\right)$ for the detector B.
 The lines are the theoretical formula Eq.~(\ref{eq:V_B_omega}). 
}
    \label{fig:spectra_low_P}
\end{figure}
%
%
%
%
%

In Fig.~\ref{fig:spectra_low_P} 
we report the experimental spectra for the case of low pressure $p^{(lo)}$ 
applying the feedback loop $(g\neq 0)$. 
The lines correspond to the theoretical fitting, whereas points correspond to experimental data.

%
%
%
%
\begin{figure}[t]
    \centering    
\includegraphics[width = 0.9 \linewidth]{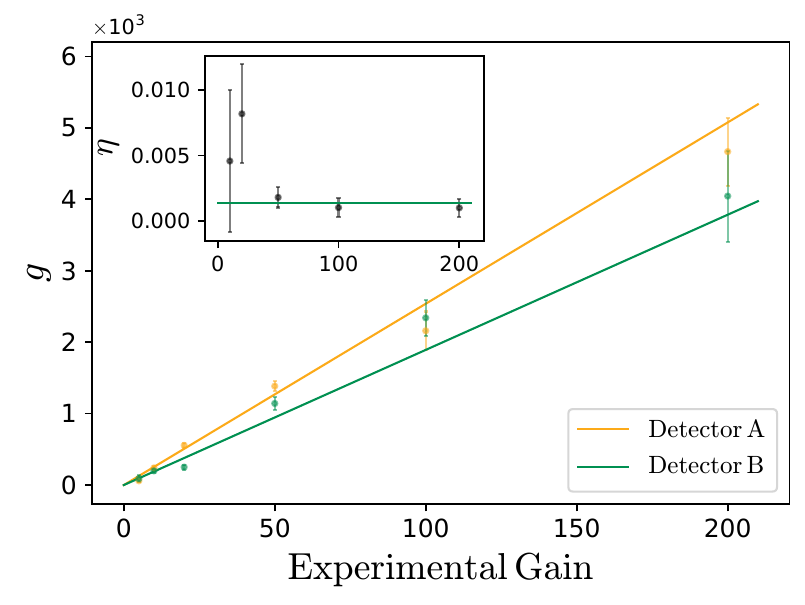}
 \caption{
For the low-pressure data, $p^{(lo)}$, the fitting parameter $g$ (scaled gain), together with its associated error, is shown for the two detectors across the different data sets acquired at varying experimental gains.
The inset shows the results for the fitting parameter $\eta$, representing the cross-talk coefficient, for detector $B$.
}
    \label{fig:g_low_P}
\end{figure}
%
%
%
%
%

In Fig.~\ref{fig:spectra_low_P}, the reduction in the area under the resonance peak indicates a lower effective temperature of the Brownian angular motion.
Compared to the high-pressure data, Fig.~\subref{fig:spectra_low_P}{a} shows that for detector $A$, included in the feedback loop, the squashing effect 
(namely the depletion below the noise level)  
becomes pronounced at high gain values due to the damping coefficient being reduced by two orders of magnitude.
Meanwhile, Fig.~\subref{fig:spectra_low_P}{b} shows that for detector $B$, which is not part of the feedback loop, the crosstalk effect (namely the asymmetry of the curve) is more pronounced at low pressure.

In Fig.~\ref{fig:g_low_P}, we report the fitted parameters $g$ at low pressure for the different data sets at varying experimental gains, for both detectors.
The first observation is that the ratio between the scaled gain $g$ curve at low pressure and the corresponding $g$ curve at high pressure (Fig.~\ref{fig:g_high_P}) is approximately a factor of $200$.
Within statistical uncertainty, this ratio coincides with the ratio between the damping coefficient at high pressure and that at low pressure.
This confirms the internal consistency of the analysis, as the theoretical scaled gain is expected to depend inversely on the damping coefficient (see Eq.~\ref{eq:g}).

%
%
%
%
\begin{figure}[t]
    \centering
\includegraphics[width = 0.9 \linewidth]{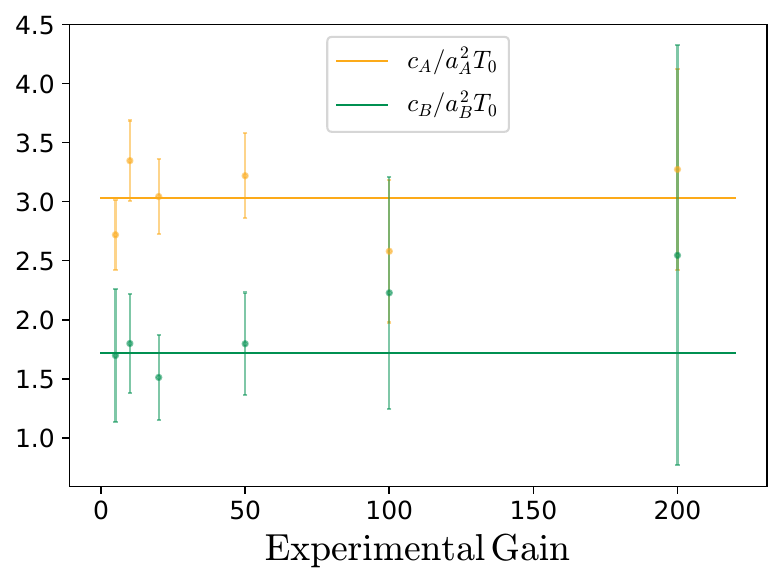}
 \caption{
The ratios  $c_A / (a_A^2 T_0)$ and $c_B/ (a_B^2 T_0)$, 
where the denominators refer to the high-pressure values 
whereas $c_A$ and $c_B$ are obtained by fitting  the low-pressure data, $p^{(lo)}$, for the different data sets at varying
experimental gains (see text). 
}
    \label{fig:g_low_P_2}
\end{figure}
%
%
%
%
%

\subsubsection{Analysis of the fitting parameters $c_A$ and $c_B$.}
In Fig.~\ref{fig:g_low_P_2} we report the results of the fitting for the parameters $c_A$ and $c_B$ 
scaled with the corresponding value at high pressure, namely 
  $c_A/ ( a_A^2 T_0)$ and $ c_B / ( a_B^2 T_0)$.

The first striking observation is that the ratios are different from one. This justifies the choice of leaving these parameters free, but we need to find a possible explanation.
A priori, upon varying only the pressure, one would expect that neither the temperature inside the cryostat nor the calibration factors should change.

However, two considerations must be taken into account.

First, to reduce the pressure in the chamber, we need to connect a pumping system and later disconnect it to minimize vibrational noise. 
In this process, there may be a slight change in the effective orientation and tilt of the chamber, which could lead to a slight change in the equilibrium angle $\theta_0$. 
Indeed, from  Fig.~\ref{fig:g_low_P_2}  
we can infer 
that the ratio of the signals in the two channels is changed.
We can estimate the angle using the fact that 
$\tilde{a}_A^2/\tilde{a}_B^2 = c_A/c_B \simeq (3.03/1.75) 
a_A^2/a_B^2 $ 
finding $\tilde{\theta}_0 \approx 26$ deg, revealing a $6$ deg shift.

Second, it is plausible that the zero-gain temperature $T$ characterizing the effective thermal motion of the levitated micromagnet does not necessarily coincide with the nominal temperature of the cryostat $T_{\textrm{0}}$.
Although we are confident that at this pressure the micromagnet is well-thermalized, an effective noise temperature higher than $4.2$ K could arise from vibrational (seismic or acoustic) noise from the laboratory environment. The latter would be negligible at high pressure, but could become significant at low pressure, where the gas damping and the corresponding thermal torque noise are two orders of magnitude smaller.
As shown in Appendix~\ref{app:zerogT}, we can estimate the zero-gain temperature $T$ at low pressure by taking into account the change in couplings with the two coils, obtaining 
$T = 7.9 \pm 1.3~\mathrm{K}$.

%
%
%
%
\section{Cooling temperature} 
\label{sec:Temperature}

In this section, we analyze the effective cooled temperature reached by using the magnetic feedback cooling presented in the previous section.
Hereafter we focus the analysis on the low-pressure data.


The average potential energy and the average kinetic energy of the Brownian angular motion are related to the power spectrum  
%
%
%
%
%
%
%
%
%
%
%
%
%
\begin{align}
E_{\mathrm{pot}}& = \frac{1}{2} \mathcal{I} \omega_0^2 \langle \theta^2 \rangle =  \frac{1}{2} \mathcal{I} \omega_0^2  \int^{+\infty}_{-\infty}\!\! \frac{d\omega}{2\pi}  
S_{\theta}\left( \omega \right) 
\label{eq:E_pot} \\
E_{\mathrm{kin}}& = \frac{1}{2} \mathcal{I} \langle \dot{\theta}^2 \rangle =  \frac{1}{2} \mathcal{I}   \int^{+\infty}_{-\infty}\!\! \frac{d\omega}{2\pi}   \omega^2
S_{\theta}\left( \omega \right)  
\label{eq:E_kin} 
\end{align}
%
%
%
%
%
%
%
%
%
%
%
%
%
from which one can compute the total average energy and the associated effective temperature  defined as 
%
%
%
%
%
%
%
%
%
%
%
%
%
\begin{equation}
k_B T_{\mathrm{eff}} = E_{\mathrm{tot}} =  E_{\mathrm{pot}} + E_{\mathrm{kin}} \, .
\label{eq:T_eff_def}
\end{equation}
%
%
%
%
%
%
%
%
%
We note that we have defined the effective temperature as being associated with the total average energy, 
since, strictly speaking, the equipartition theorem does not hold in a general feedback cooling scheme 
\cite{Genes:2008,Habibi:2016}.
However, in the regime of the experimental parameters 
and range of the gain, this asymmetry between the average kinetic and potential energy is very small and, for this reason, one can still safely associate the total energy with an effective temperature (see Appendix \ref{app:equipartition}).

For $g=0$ it is straightforward to recover the expected result, $k_B T_{\mathrm{eff}} \rightarrow k_B T$, from the spectrum 
$S_{\theta}(\omega)$ with the susceptibility $\chi_0(\omega)$. 
For $g>0$, 
it is useful to introduce the dimensionless scaled ratio corresponding to the scaled detector noise
%
%
%
%
%
%
%
%
%
%
%
%
%
\begin{align}
\Theta_{n} &
= 
\frac{\omega_0^2 \, \mathcal{I}  \,\tilde{\Gamma}_0 \,S_{\delta A}}{2   k_B T\,\tilde{a}_A^2}
\label{eq:theta_n}
\end{align}
%
%
%
%
%
%
%
%
%
%
%
%
%
where $S_{\delta A}$  and $\tilde{a}_A$  are, respectively, 
the detection noise  
and
the calibration factor 
of the detector $A$ used in the feedback loop.

We remark that $\Theta_{n}$ depends on the zero-gain temperature $T$ and on the calibration factor 
only through the combination $ \tilde{a}_A^2 T$.
It is possible to show (see Appendix \ref{app:theory}) that the  quantity $T_{\mathrm{eff}}/T$ is only a function of the scaled gain $g$, the scaled noise  $\Theta_n$ and
the ratios $\omega_c/\omega_0$ and $\tilde{\Gamma}_0/\omega_0$.
Proceeding in this way, one can extract the temperature reduction and thus quantify the cooling induced by the magnetic feedback, without requiring precise knowledge of the zero-gain temperature or the coupling to the detector.

\begin{figure}[t]
    \centering
\includegraphics[width = 0.9 \linewidth]{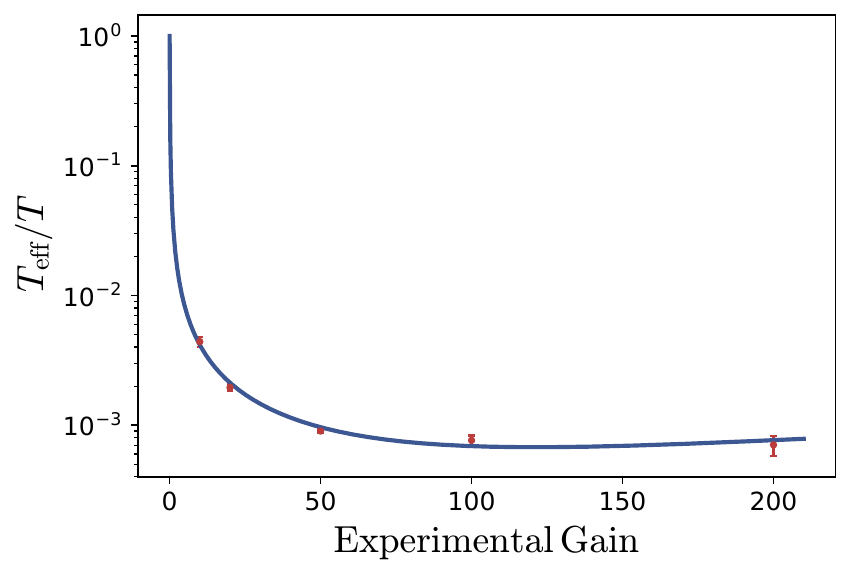}
 \caption{The  effective temperature scaled with the zero-gain temperature, $T_{\mathrm{eff}} /T$, 
 as a function of the experimental gain $G$.
 The points are obtained as explained in the text.
The solid line is computed using 
$g = 25.4 \,G$, 
corresponding to the linear fit in 
Fig.~\ref{fig:g_low_P}, and the 
coefficient 
$c_A =3.03\,a_A^2 T_0$, as obtained by the fitting in  Fig.~\ref{fig:g_low_P_2}.
}
    \label{fig:cooling}
\end{figure}

We plot $T_{\mathrm{eff}}/T$ in Fig.~\ref{fig:cooling} as a function of the experimental feedback gain $G$.
Each point of $T_{\mathrm{eff}}/T$ is 
computed using the theoretical scaled gain $g$, as obtained from 
the fitting procedure for each data set, Fig.~\ref{fig:g_low_P}. 
Similarly, the scaled noise 
is computed using the coefficient 
$c_A = \tilde{a}_A^2 T$ obtained from the fit in 
Fig.~\ref{fig:g_low_P_2}.
We apply standard error propagation for the vertical 
error bars.

As is well known from the general theory of feedback cooling, 
the curve exhibits a minimum: while increasing the gain initially enhances the cooling efficiency, 
a further increase leads to the reinjection of detection noise into the microparticle, 
resulting in heating rather than cooling \cite{Poot:2012}. 
In our experiment, we operated near the optimal feedback gain, corresponding to the maximum achievable temperature reduction set by the detector noise.

Figure~\ref{fig:cooling} shows that the angular motion is cooled by three orders of magnitude.
Combining this result with the zero-gain temperature estimated above allows us to determine the lowest temperature reached.
In particular, for the data point at $G = 100$, which lies close to the minimum, we obtain $T_{\mathrm{eff}} = 6 \pm 1~\mathrm{mK}$.

%
%
%
%
%
%
%
%
%

%
%
%
%
\section{Discussion and Perspectives} 
\label{sec:discussion} 

\subsection{Discussion}

As summarized in Fig. \ref{fig:cooling}, we have experimentally realized the feedback-cooling of a librational mode of a levitated micromagnet down to $6 \pm 1~\mathrm{mK}$ starting from a bath temperature $T_0=4.2$ K.
This result is comparable to the strongest cooling obtained so far with a Meissner-levitated ferromagnet \cite{Uitenbroek:2026}.

Compared with previous implementations \cite{Uitenbroek:2026, Timberlake:2024}, the unique feature of our setup is the feedback scheme based on a purely magnetic actuation. Unlike piezoelectric schemes, magnetic feedback is highly efficient on librational modes, generates negligible heating and can be easily integrated into a miniaturized setup on-chip \cite{wieczorek2023chip}. Furthermore, it can be optimized on specific modes by tailoring the coil geometry and arrangement. In our highly symmetric setup we have demonstrated a strong suppression of cross-talk between feedback coil and the pick-up coil A, a feature which avoids spurious instabilities in the closed loop system.

A further novel aspect of our scheme is the use of a double detection scheme. In the present work only one channel is used for feedback, while the second channel is used only to monitor the cooling under squashing conditions. 
%
The cross-talk issue can be prevented by having the pick-up coils $A$ and $B$ overlapped.
%
However, one may devise schemes in which the information from two or more channels is combined to provide feedback on different modes. In this case, nonoverlapping pick-up coils with coupling optimized on different modes are a potentially better choice.

In future implementations, some technical aspects need to be improved. We have observed an unwanted change in the coupling factors to the pick-up coils when changing pressure. This is compatible with a change of the equilibrium orientation. Currently, we do not have a full picture of the possible mechanisms leading to this effect. A slight increase of the no-damping temperature is also observed at low pressure, likely pointing to small torque noise contributions from seismic noise or vibrations generated inside the cryostat.

Aside from technical noise, the minimal effective temperature achievable by feedback-cooling is clearly limited by detection noise. This is apparent from Fig. \ref{fig:spectra_low_P} which shows that we have entered the squashing regime, meaning that we have achieved the ultimate cooling available with the experimental detection noise. Concretely, the poor detection noise can be largely attributed to the relatively weak magnetomechanical coupling. In the next section, we will discuss the ultimate limitations from fundamental detection noise at the quantum limit and thermal noise, outlining a pathway towards ground-state cooling.

Finally, we have developed a full theoretical framework which takes into account all specific features of the feedback scheme, slightly different from the standard cold damping. The full theory allows us to reach a comprehensive understanding of all features of the spectra under feedback-cooling, beyond the usual narrowband approximation.

\subsection{Towards ground-state cooling}

In this section, we outline the path towards cooling a librational mode of a levitated magnet to the ground state, thereby entering the quantum regime of angular motion. 
In order to cool a mechanical oscillator to the ground state by measurement and feedback, one needs a quantum-limited detector and the dynamics of the resonator has to be dominated by quantum back-action rather than by environment-induced thermal noise \cite{Rossi_Mason_Chen_Tsaturyan_Schliesser_2018,Tebbenjohanns:2021,Magrini:2021,Kamba:2023}. In the experiment described in this work, we are still far from these conditions. 

A SQUID can in principle reach the quantum limit as a magnetic flux detector \cite{koch1980quantum,falferi2008}, meaning that the product of imprecision and back-action noise spectral densities approaches the lower bound imposed by the Heisenberg uncertainty limit \cite{braginsky1975,caves1982}, expressed mathematically by
\begin{equation}
  S_\Phi^{\imp}S_{J}^{\BA} \geq \hbar^2/4   \label{eq:quantumlimit}
\end{equation}
where $S_\Phi^{\imp}$ and $S_{J}^{\BA}$ are the two-sided spectral densities of the imprecision flux noise and of the back-action circulating current noise.
When using a SQUID to detect the angular displacement $\theta$ of a ferromagnet, with magnetomechanical coupling $\Phi_\theta = d\Phi/d\theta$, the same relation applies to the product of the angular imprecision spectral density $S_\theta^{\imp}=S_\Phi^{\imp}/\Phi_\theta^2$ and back-action torque spectral density $S_\tau^{\BA}= \Phi_\theta^2 S_J^{\BA}$.

In the following we will assume we have a quantum-limited SQUID. We will discuss later the feasibility and the technical limitations. To operate in a regime where the quantum back-action becomes apparent, we need to increase the magnetomechanical coupling by many orders of magnitude compared to the current setup. Let us estimate the maximum realistically achievable  coupling. We model the magnet as a sphere with radius $R$, magnetization $M$ and dipole moment $\mu=4\pi/3 R^3 M$, and the SQUID as a circular loop of radius $r_s > R$ where $R$ is the magnet radius. To maximize the coupling we place the magnet in the SQUID loop center with equilibrium orientation in-plane and a librational mode corresponding to a rotation towards the loop axis. Under these conditions the magnetomechanical coupling reads
\begin{equation}
  \Phi_\theta = \frac{\mu_0 \mu}{2 r_s}.  \label{eq:phitheta}
\end{equation}
Note that the circular loop can be the SQUID itself but could be as well a superconducting pick-up loop coupled to a remote SQUID, as in the current experiment. Under optimal impedance matching the loss of coupling of a pick-up coil configuration can be made small \cite{clarke1979magnetometer}. 

Let us consider a realistic configuration with $R=30$ $\mu$m and magnetization $\mu_0 M=0.85$ T, similar to our current magnet, leading to a magnetic dipole $\mu=7.6\times 10^{-8}$ J/T and moment of inertia $\mathcal{I}=3 \times 10^{-19}$ kg\,m$^2$, and a SQUID radius $r_s = 10 R=300$ $\mu$m, which makes the field at the loop wire much smaller than the critical fields. By means of Eq. (\ref{eq:phitheta}) we obtain a coupling $\Phi_\theta = 1.6 \times 10^{-10}$ Wb/rad. The optimal noise of a dc-SQUID close to the quantum limit is $S_\Phi^{\imp}\approx \hbar L$ \cite{tescheclarke1977,koch1980quantum}. The inductance of a superconducting circular loop is~\cite{clem2003}
\begin{equation}
L=\mu_0 r_s [\ln(8r_s/w)-1.75 ] \label{eq:L}
\end{equation}
where we assume a wire radius $w=1 \, \mu$m (the actual value is not so important due to the weak logarithmic dependence). This yields an imprecision angular spectral density $S_\theta^{\imp}=9 \times 10^{-24}$ rad$^2$/Hz and a quantum back-action spectral density $S_\tau^{\BA}=7 \times 10^{-47}$ N$^2$m$^2$/Hz.

The latter should be compared with the environment-induced thermal noise, expressed by $S^{\tau}_{\mathrm{th}}= 2 k_B T \Gamma_0 \,  \mathcal{I}$. With $T=4.2$ K and $\Gamma_0=10^{-6}$ Hz, we obtain $S^{\tau}_{\mathrm{th}}=3 \times 10^{-47}$ N$^2$m$^2$/Hz, less than half the quantum back-action. This shows that a magnetically levitated particle can be in principle cooled to the ground state even starting from $T=4.2$ K, provided that $\Gamma_0$ is sufficiently low and a sufficiently strong coupling to a quantum limited SQUID detector can be realized. Note that even larger ratios between quantum back-action and thermal noise can be achieved by reducing the temperature or further increasing the coupling by reducing $r_s$. For instance, with $T=0.5$ K, $\Gamma_0=10^{-6}$ Hz and $r_s = 6R$ the thermal noise becomes about $1 \%$ of the quantum back-action.
For a system with the parameters discussed above, starting from an initial temperature of $T = 4.2~\mathrm{K}$, the standard cold-damping model predicts a minimum achievable temperature of $T_{\min} = 2~\mathrm{nK}$ when the backaction noise is included as a constant contribution. This value corresponds to reaching the ground state of the system.

\subsection{Experimental challenges}

We are now going to discuss the experimental challenges and potential limitations of the above analysis. First, let us consider the feasibility of a quantum-limited SQUID. Conventional resistively shunted dc-SQUIDs, such as those used in this work, are a well-established and robust technology. In particular, they can be routinely operated within feedback loops with very large dynamics. However, they have approached \cite{falferi2008} but never fully achieved the quantum limit in terms of Eq. (\ref{eq:quantumlimit}). Filling this gap is known to be hard, because of the hot-electron effect in the shunt resistors \cite{wellstood1994hotelectron}. On the other hand, SQUID-based microwave cavities can be operated as sensitive quantum limited magnetic flux detectors \cite{zoepfl2020,schmidt2024}. The downside is that, unlike in dc SQUIDs, the dynamic range is very poor even under feedback operation, a feature which is needed to deal with large background signals. Another practical technical issue for operation down to 100 Hz, common to dc-SQUIDs or SQUID-based cavities, is given by excess 1/f flux noise, which is thought to arise from defects and impurities in superconducting thin films \cite{wellstood1987unosueffe}. We point out that all these issues are in principle technical, implying that there are no fundamental limitations preventing a SQUID from operating as a quantum-limited detector of the angular rotation of a ferromagnet down to 100 Hz. The practical realization of a fully quantum-limited SQUID at $1-100$ Hz though still appears challenging.

The next challenge is to achieve a sufficiently large coupling. In current experiments the pick-up coil is made of a wire wound by hand on a plastic support machined in a mechanical workshop. The distance between coils and magnets is of the order of $2$ mm. Machining a coil with the radius $r_s=10R\approx 300$ $\mu$m  and aligning it in the setup with the required accuracy appears unfeasible. However, the realization of such a coil is totally feasible on-chip, as already demonstrated in a similar type of levitation \cite{wieczorek2023chip}. We are currently investigating such an approach.

Finally, let us consider thermal noise. The current experiment is operated at $T=4.2$ K, and values of $\Gamma_0$ down to $10^{-4}-10^{-5}$ Hz have been achieved in translational \cite{Fuchs:2024,Vinante:2020PRApplied,Vinante:2020PRApplied}. 
and librational modes \cite{Vinante:2020PRApplied}. In the spinning mode with a magnet similar to the one considered in this work, a record low damping $\Gamma_0= 4.7\times 10^{-7}$ Hz has been reported \cite{Jose:2026}. Thus, the required numbers appear within reach. At present, the biggest issue is the suppression of ambient vibrations, a particularly challenging task in cryostats at frequencies below 1 kHz \cite{Fuchs:2024}. For a magnet with parameters as in the previous estimations, and for a pure librational mode, the required residual angular acceleration should be much lower than $(S_\tau^{\BA})^{1/2}/\mathcal{I} \approx 10^{-5}$ rad/s$^2$/Hz$^{1/2}$, which appears  feasible.

\section{Conclusions} 
\label{sec:conclusions}

To summarize, 
we demonstrated magnetic feedback cooling of a levitated micromagnet in a superconducting Meissner trap using integrated magnetic coils for detection and feedback.

A dual detection scheme is employed, enabling independent monitoring of the Brownian motion and revealing effects such as squashing and cross-talk.

Feedback cooling is applied in both high- and low-pressure regimes, with different damping conditions. 

By analyzing the power spectral density within a theoretical model of the feedback loop, we find that magnetic feedback cooling reduces the effective temperature of the levitated micromagnet by approximately three orders of magnitude, for a low-frequency mechanical mode ($f_0 \sim 100$  Hz) with a mass in the nanogram range.
Assuming a zero-gain temperature of $T = 7.9 \pm 1.3~\mathrm{K}$, we obtain a minimum effective temperature of $T_{\mathrm{eff}} = 6 \pm 1~\mathrm{mK}$ for the librational motion of the micromagnet.

Finally, we have discussed a realistic scenario for potential improvements of the system that would enable the cooling of a librational mode to the ground state, corresponding to the nanokelvin regime, thereby opening the door to quantum experiments with massive magnetically levitated systems.

%

\begin{acknowledgments}
%
%
%
%
P.O., G.R. and  A.V. acknowledge financial support from the Q@TN Initiative 
Provincia Autonoma di Trento (PAT). 
A.V. and A.M. acknowledge support from the Italian Ministry for University and Research within the Italy-Singapore Scientific and Technological Cooperation Agreement 2023-2025. 
A.V. acknowledges support from the QuantERA II Programme (project LEMAQUME) that has received funding from the European Union’s Horizon 2020 research and innovation programme under Grant Agreement No 101017733.
P.O. and G.R. also acknowledge direct support from the Provincia Autonoma di Trento (PAT) for the BEC Center.  

\end{acknowledgments}

\appendix

%
%
%
\section{Additional measurements}
\label{app:ringdown_cross_talking}

%
%
%
%

\subsection{Ringdown measurement at low pressure}

While in the high-pressure curves the damping can be readily estimated from the spectra even at zero gain, for the low-pressure case the zero-gain damping is $\tilde{\Gamma}_0/(2\pi) < 1$ mHz, a value which makes a spectral estimation highly inconvenient. In this regime, it is much more practical to estimate the damping by measuring the ringdown decay rate with a lock-in amplifier tuned on resonance. 

\begin{figure}[t]
    \centering
\includegraphics[width = 0.9 \linewidth]{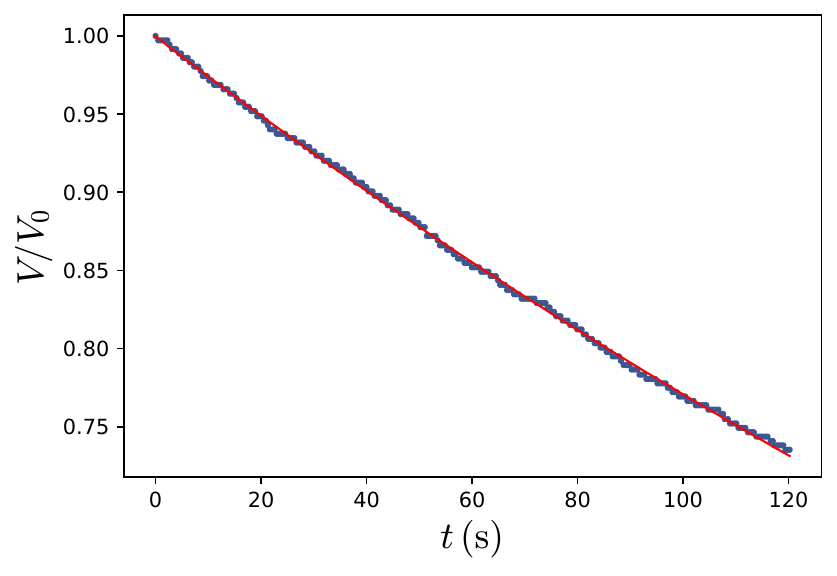}
 \caption{Ringdown measurement for the case of low pressure and no feedback gain. The measurement is used to infer the intrinsic damping at low pressure $\tilde{\Gamma}_0$. The solid line is the best exponential fit to the data.
}
    \label{fig:ringdown}
\end{figure}

An amplitude ringdown measurement for the low-pressure, zero-gain case is shown in Fig.~\ref{fig:ringdown}. Here, the librational mode was excited by direct driving at resonance with one of the coils. 
The solid red line is an exponential fit with the function $\theta(t)=\theta_0 \exp{(-\tilde{\Gamma}_0 t/2 )}$, which describes the envelope of the free amplitude decay according to Eq.~\eqref{eq:Langevin}. The fit yields $\tilde{\Gamma}_0 = 5.20 \times 10^{-3} s^{-1}$. 
The uncertainty is much lower than $1 \%$ and is practically not significant compared to other sources of uncertainty in the experiments.

%
%
%
%

\subsection{Cross-talk}

Direct magnetic feedback is a convenient method in our magnetic setup, but comes with the drawback of unavoidable direct crosstalk between the feedback and the detection coils. We partially overcome this issue by realizing our setup with eight-shaped coils wound on the same support but with two possible orthogonal orientations. 
As shown in Fig. \ref{fig:system}, the two detection coils A, B are coplanar but rotated by $\pi/2$ with respect to each other, which gives a different sensitivity to $\theta$ motion. The feedback coil C lies on a different parallel plane and is parallel to B and orthogonal to A. We thus expect the coupling to A to be highly suppressed due to symmetry. 

We can directly measure the coupling of C to A and B by sending a current to the coil C and measuring the output of the SQUIDs A and B. The measured ratio between the output of SQUID A and SQUID B is $2.2 \times 10^{-3}$. We can estimate the absolute coupling between the coils, taking into account the nominal values of the SQUID input coil inductance $L_i=1.8 \, \mu$H, the input coil mutual inductance $M_i = 10$ nH (both parameters are nominally identical for the two SQUIDs), and the measured flux-to-voltage electronics gain $V_\Phi = 0.31$ $V/\Phi_0$ for both channels. We obtain $M_{CA}=22.8$ pH and $M_{CB}=10.4$ nH. 

The coupling from C to A is thus practically negligible, while the coupling from C to B leads to the undesired asymmetry of the spectrum at large feedback gain.

%
%
%
\section{Estimation of zero-gain temperature}
\label{app:zerogT}
%
%
%
%
As anticipated in the main text, when considering the low-pressure regime, we cannot rule out the possibility that additional noise sources contribute to the effective thermal noise, thereby increasing the zero-gain temperature with respect to that in the high-pressure regime.

To estimate this variation, we express the calibration factors in terms of the conversion parameter $a$ and the angle $\theta_0$ of the magnetic moment with respect to the axis of coil A
\begin{equation}
	a_A = a \sin \theta_0 \, , \qquad a_B = a \cos \theta_0 \, , 
	\label{eqn:zerogT.1}
\end{equation}
At high pressure this yields an estimate of the angle $\theta_0 = \arctan(a_A/a_B) = 20.2 \pm 1.0~\mathrm{deg}$.

The same relations hold at low pressure, where we allow for a different angle $\tilde{\theta}_0$. However, we cannot estimate the calibration factors $\tilde{a}_A$ and $\tilde{a}_B$ directly, as the fit only provides access to the products $c_A = \tilde{a}_A^2 T$ and $c_B = \tilde{a}_B^2 T$. Evaluating the ratios between these parameters and the corresponding high-pressure ones,
\begin{equation}
	X_A = \frac{\tilde{a}_A^2 T}{a_A^2 T_0} \, , \qquad X_B = \frac{\tilde{a}_B^2 T}{a_B^2 T_0} \, , 
	\label{eqn:zerogT.2}
\end{equation}
and taking their ratio
\begin{equation}
	\frac{X_A}{X_B} = \frac{\tilde{a}_A^2}{\tilde{a}_B^2} \frac{a_B^2}{a_A^2} = \left( \frac{\tan\tilde{\theta}_0}{\tan\theta_0} \right)^2 \, , 
	\label{eqn:zerogT.3}
\end{equation}
we obtain the new angle $\tilde{\theta}_0 = 26.3 \pm 1.4~\mathrm{deg}$. Knowing both angles, we can invert one of the relations in Eq.~\eqref{eqn:zerogT.2} to find
\begin{equation}
	T = T_0 X_A \left( \frac{\sin\theta_0}{\sin\tilde{\theta}_0} \right)^2 = 7.9 \pm 1.3~\mathrm{K} \, .
	\label{eqn:zerogT.4}
\end{equation}

%
%
%
\section{Some details of the theoretical calculations}
\label{app:theory}
%
%
%
%

Here we present some main steps that we used to obtain the equations reported 
in the main text.

\subsection{Scaling} 

As stated in the main text, we describe the dynamics of the librational motion, parameterized by the angle $\theta(t)$ around the equilibrium position, using the  Langevin equation Eq.~\eqref{eq:Langevin}.
The feedback torque $\tau_{\fb}(t)$, in terms of the angular variable (i.e., after the voltage-to-angle conversion), is given by Eq.~\eqref{eq:fb_function} 
with $\delta\theta_A(t)$  the angular noise associated with the output of detector A.
From Eq.~\eqref{eq:Langevin}, we obtain the equation in the frequency domain
\begin{multline}
	\big(\omega_0^2 - \omega^2  - i \omega \Gamma_0 \big) \theta(\omega) = \\
		= \delta \tau_{\th}(\omega) + K_{\fb}(\omega) \big( \theta(\omega) + \delta\theta_A(\omega) \big) \; . 
\end{multline}
From this equation, we can readily identify the response function Eq.~\eqref{eq:response_f} 
which allows us to express the angular position in the frequency domain as Eq.~\eqref{eq:Langevin_omega}.
To evaluate the power spectrum, we exploit the Wiener-Khinchin theorem
\begin{equation}
	S_{\theta}(\omega) = \int_{-\infty}^{+\infty} \!\!\!\!\!\! dt \, e^{ i\omega t} \langle \theta(t) \theta^*(0) \rangle \; .
\end{equation}
Thus, we must determine the correlation $\langle \theta(t) \theta^*(0)\rangle$ of the angular variable, which easily follows from the correlation of its Fourier transform, $\langle \theta(\omega_1)\theta^*(\omega_2)\rangle$. 
Proceeding in this way, one obtains Eq.~\eqref{eq:S_theta} which in the scaled form reads
\begin{equation}
	S_{\theta}(\omega) = \frac{2\Gamma_0 k_B T}{\mathcal{I}} {\left| \chi(\omega) \right|}^2 \left( 1 + g^2 \frac{\omega_0^2}{\omega_c^2 + \omega^2} \Theta_{n, A} \right) \, ,
\label{eq:S_theta_expl}
\end{equation}
with $\Theta_{n, A} $ the scaled 
detection noise Eq.~\eqref{eq:theta_n}.
Then  
we can integrate the power spectrum of the angle to evaluate the mean-square angular fluctuations Eq.~\eqref{eq:E_pot}  and 
integrate the power spectrum of the mean-square value of the angular velocity,  $S_{\dot{\theta}}(\omega)=\omega^2 \, S_{\theta}(\omega)$, 
 to evaluate the mean-square angular velocity fluctuations, Eq.~\eqref{eq:E_kin}.
Using these  results, we can compute the total average energy associated with the system, Eq.~\eqref{eq:T_eff_def}, which we can cast 
in terms of rescaled parameters
\begin{align}
	& \frac{E_{\mathrm{tot}}}{E^{(g=0)}_{\mathrm{tot}}} = \notag \\
	& = \frac{\gamma_0}{\pi} \int_{0}^{+\infty} \!\!\!\! dz \,  \frac{\left( 1 + z^2 \right) \left( 1 + g^2 f_c(z) \Theta_{n, A} \right)}{ \big( 1 - z^2 + g z_c \gamma_0 f_c(z)\big)^2 + z^2 \gamma_0^2 \big( 1 + g f_c(z) \big)^2} \; ,
\label{eq:XXX-appendix}
\end{align}	
showing that the ratio is purely a function of $g$ and of the scaled parameters
\begin{equation}
	\gamma_0 = \frac{\Gamma_0}{\omega_0} \, , \quad  z_c = \frac{\omega_c}{\omega_0} \; , \qquad \text{ and } \, f_c(z) = \frac{1}{z_c^2 + z^2} \; ,
\label{eq:parameters_rescaled}
\end{equation}	
as stated in the main text.
We used Eq.~\eqref{eq:XXX-appendix} for Fig.~\ref{fig:cooling}, using the parameters extracted by the fitting procedure of the power spectra.

\subsection{Power spectra}

We now discuss the spectra associated with the measured angular positions in Eq.~\eqref{eq:V_A} and Eq.~\eqref{eq:V_B}.

For evaluating the spectra of the measured quantities, we must compute the various correlations.
This is straightforward since the thermal noise is uncorrelated with the detectors' noise.
However, we remark that, because the signal from coil A is reinjected into the system, we have
\begin{equation}
	\langle \theta(\omega_1) \delta\theta^*_A(\omega_2) \rangle \ne 0 \, , \qquad \langle \theta(\omega_1) \delta\theta^*_B(\omega_2) \rangle = 0 \, . 
\label{eq:corr_noise_AppB}
\end{equation}	
In other words, the stochastic dynamics of $\theta(t)$ is affected not only by the thermal noise but also by the noise of detector A used in the feedback loop, which yields a non-zero correlation in Eq.~\eqref{eq:corr_noise_AppB}.
This distinction accounts for the different results shown in Eq.~\eqref{eq:V_A_omega} and Eq.~\eqref{eq:V_B_omega}.

%
%
\vspace{1em}
\paragraph*{Detector A}
For the signal read by detector A, the correlation is given by
\begin{multline}
	\langle V_A(\omega_1)V_A^*(\omega_2) \rangle = \\
		= a_A^2 \Big[ \langle \theta(\omega_1)\theta^*(\omega_2) \rangle + \langle \delta\theta_A(\omega_1) \delta \theta^*_A(\omega_2) \rangle \\
			+ \Big( \langle \theta(\omega_1)\delta\theta^*_A(\omega_2) \rangle + \langle \delta \theta_A(\omega_1)\theta^*(\omega_2) \rangle \Big) \Big] \, , 
\end{multline}
yielding the result presented in~\eqref{eq:V_A_omega}.
%
%
\vspace{1em}
\paragraph*{Detector B}

For detector B, we take into account the cross-talk contribution, which arises from the mutual inductance between coil C, which injects the feedback into the system, and the readout coil B. 

Since the signal in coil C is proportional to the one read by detector A, we assume the cross-talk term depends on $V_A(\omega)$. Its filtering function is identical to that of the feedback~\eqref{ed:ct}, apart from a constant factor that accounts for coefficients of various origins (see main text).
This produces  the result 
\begin{align}
	& \langle V_B(\omega_1)V_B^*(\omega_2) \rangle = \notag\\
		& \, = a_B^2 \Big[ \left( 1 + K_{\ct}(\omega_1) \right)\left( 1 + K_{\ct}(\omega_2) \right)^* \langle \theta(\omega_1)\theta^*(\omega_2) \rangle  \notag \\
		& \quad + K_{\ct}(\omega_1) K_{\ct}^*(\omega_2)\langle \delta\theta_A(\omega_1) \delta \theta^*_A(\omega_2) \rangle  \notag \\
		& \qquad + \left( 1 + K_{\ct}(\omega_1) \right) K_{\ct}^*(\omega_2) \langle \theta(\omega_1) \delta\theta^*_A(\omega_2) \rangle  \notag \\
		& \qquad +  K_{\ct}(\omega_1) \left(1 + K_{\ct}(\omega_2) \right)^* \langle \delta\theta_A(\omega_1) \theta^*(\omega_2) \rangle  \notag \\
		& \hspace{6em} + \langle \delta\theta_B(\omega_1) \delta \theta^*_B(\omega_2) \rangle \Big]
\end{align}
which leads to the result shown in~\eqref{eq:V_B_omega}.

%
%
\subsection{Violation of the equipartition}
\label{app:equipartition}

From Eq.~\eqref{eq:Langevin} and Eq.~\eqref{eq:fb_function}, we observe that the system is affected by two noises, the thermal one and the detector’s noise, which are uncorrelated and whose fluctuation strengths differ and cannot be identified with a single temperature.
In other words, the system is coupled to two baths at different temperatures.
In such a nonequilibrium regime, the equipartition of the kinetic and potential energy does not have to hold \cite{Genes:2008,Habibi:2016}.

In the standard cold damping method \cite{Poot:2012,Poggio:2007}, the effective 
temperature is given by the integration of the spectrum associated 
with the position $x$ of the resonator  \cite{Poot:2012}
\begin{equation}
\frac{1}{2} m\omega_0^2 \langle x^2 \rangle
= \frac{1}{2} k_B \frac{T_0+g^2 T_n}{1+g} \, , 
\end{equation}
where $T_n$ is the temperature associated with the detector noise.

Remarkably,  
the Langevin equation with the feedback scheme given by the cold damping method \cite{Poggio:2007}, in which one considers the time derivative of the detector’s noise (the latter being delta-correlated in time), produces a power spectrum of the velocity that is divergent
unless the detector bandwidth is explicitly included in the calculation.
Such a bandwidth cuts off the high-frequency divergence, but it leads
to an average velocity that explicitly, and potentially weakly, depends on this cutoff, contrary to the average square of the position, which is well defined even in the absence of the cutoff.
This clearly shows that equipartition is violated.

In our case, the filtering function used prevents the divergence in the kinetic energy, as it effectively acts as a bandwidth filter for the injected noise (i.e., it is equivalent to considering a finite bandwidth for the injected noise).
Nevertheless, the asymmetry between the potential and kinetic energy remains small for not-so-large gain, in the regime of parameters involved in our experiment.

\begin{figure}[b]
    \centering
\includegraphics[width =0.9  \linewidth]{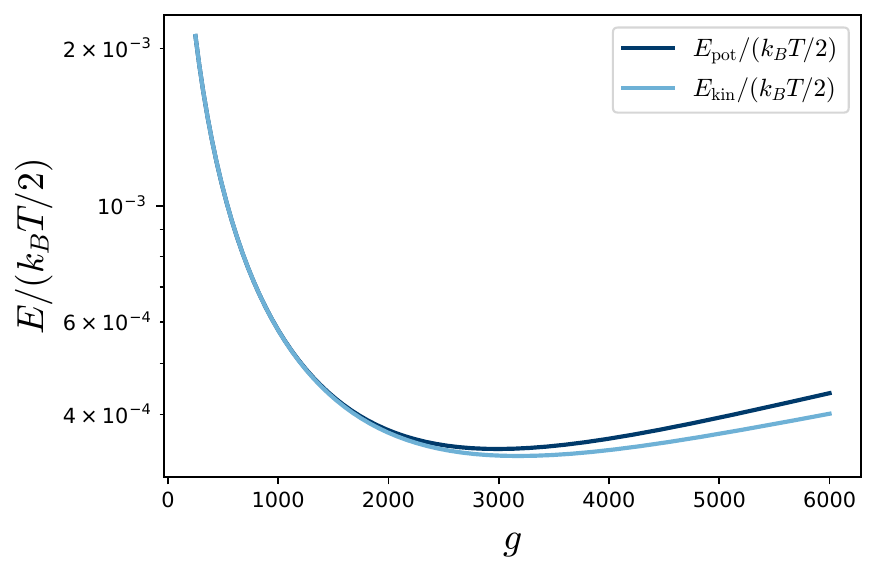}
 \caption{
 The average potential energy and the average kinetic energy as a function of the feedback gain $g$. We use the values 
 as extracted from the experimental data, namely $\Theta_n = 1.06 \cdot 10^{-7} \equiv \Theta_{n, A}$, $\Gamma_0/\omega_0= 6.1 \cdot 10^{-6}$ and $\omega_c/\omega_0= 0.23$.
}
    \label{fig:equi}
\end{figure}

In Fig.~\ref{fig:equi}, we show the average potential and kinetic energies as a function of the feedback gain $g$, using the value of $\Theta_n$ obtained from the experimental data for the in-loop detector (i.e., detector $A$), namely $\Theta_n = 1.06 \times 10^{-7}$.
We observe a clear breakdown of the equipartition theorem, as $\bar{E}_{\text{pot}} \neq \bar{E}_{\text{kin}}$ at large $g$, corresponding to the experimental points characterized by a large uncertainty (see Fig.~\ref{fig:g_low_P}).

%
%
%
%
%
%
%
%
%

%


\bibliography{references}

\end{document}